\documentclass[twocolumn]{aastex631}
\usepackage{import}

\usepackage{mathtools}
\usepackage{chngcntr}
\usepackage{bm}
\usepackage{amsmath}
\usepackage{amssymb}
\usepackage{footnote}
\usepackage[caption=false]{subfig}
\usepackage{float}
\usepackage{tabulary,booktabs}
\usepackage{multirow}
\usepackage{yhmath}
\usepackage{graphicx}
\usepackage{csquotes}
\usepackage{mwe}

\shorttitle{Star Formation in Dwarf Galaxies}
\shortauthors{Sattari et al.}
\begin{document}
\title{\large {Optical Spectroscopy of Dwarf Galaxies at $z\sim 0.15$ in the COSMOS Field: Star Formation and Dust Properties}}

\correspondingauthor{Zahra Sattari}
\email{zsattari@ipac.caltech.edu}

\author[0000-0002-0364-1159]{Zahra Sattari}
\affiliation{IPAC, California Institute of Technology, 1200 E. California Blvd, Pasadena, CA 91125, USA}
\affiliation{The Observatories of the Carnegie Institution for Science, 813 Santa Barbara St., Pasadena, CA 91101, USA}
\affiliation{Department of Physics and Astronomy, University of California, Riverside, 900 University Ave, Riverside, CA 92521, USA}

\author[0000-0003-4727-4327]{Daniel D. Kelson}
\affiliation{The Observatories of the Carnegie Institution for Science, 813 Santa Barbara St., Pasadena, CA 91101, USA}

\author[0000-0001-5846-4404]{Bahram Mobasher}
\affiliation{Department of Physics and Astronomy, University of California, Riverside, 900 University Ave, Riverside, CA 92521, USA}

\author[0000-0003-3691-937X]{Nima Chartab}
\affiliation{IPAC, California Institute of Technology, 1200 E. California Blvd, Pasadena, CA 91125, USA}

\author[0000-0001-7166-6035]{Vihang Mehta}
\affiliation{IPAC, California Institute of Technology, 1200 E. California Blvd, Pasadena, CA 91125, USA}

\author[0000-0002-7064-5424]{Harry I. Teplitz}
\affiliation{IPAC, California Institute of Technology, 1200 E. California Blvd, Pasadena, CA 91125, USA}

\author[0000-0003-3350-9869]{Shannon G. Patel}
\affiliation{The Observatories of the Carnegie Institution for Science, 813 Santa Barbara St., Pasadena, CA 91101, USA}

\begin{abstract}

We present a spectroscopic study of low-mass galaxies (LMGs;$10^8\leq\rm M_*/M_\odot\leq10^9$) at $z\sim0.15$ in COSMOS field, and compare it to a control sample of intermediate-mass galaxies (IMGs;$10^9\leq\rm M_*/M_\odot\leq10^{10}$) at $z\sim0.35$. We examine their star formation rates (SFRs), dust attenuation properties, and the relationship between nebular and stellar reddening. For both samples, SFRs derived from H$\alpha$ are strongly correlated with SFRs from fitting simple star formation histories (SFHs) to the galaxies’ spectral energy distributions. In fitting a joint SFR–$\rm M_*$ relation, we obtain a slope of $\rm {\Delta log(SFR_{H\alpha})}/{\Delta log(M_*/M_\odot)}=1.01\pm0.03$, indicating that fair ensembles of SFHs for galaxies at these stellar masses are well-described by scale-free, self-similar forms. We also examine their dust attenuation properties and the relationship between nebular and stellar reddening, exploring how these quantities vary with stellar mass and specific SFR (sSFR). Nebular attenuation increases with stellar mass for IMGs but is lower and less mass-dependent in LMGs, consistent with their reduced dust content. In all cases, stellar continuum attenuation is lower than nebular attenuation, as expected from the two-component dust model. The nebular-to-stellar color excess ratio in both samples is consistent with the canonical factor of 2.27. The ratio is mass-independent, but rises with sSFR in IMGs and remains constant in LMGs. These results suggest that in LMGs, efficient dispersal of birth clouds keeps the differential attenuation approximately constant across sSFR. Thus, although LMGs follow the same global SFR–$\rm M_*$ scaling as massive galaxies, their lower dust content and feedback-maintained ISM produce distinct attenuation behavior relative to IMGs.

\end{abstract}

\section{Introduction}\label{Intro}

Low-mass galaxies (LMGs; $\rm M_\ast \lesssim 10^9\,M_\odot$) are the most numerous systems in the universe. In the early stages of structure formation, galaxies of such masses provided the fundamental building blocks from which larger systems assembled \citep{White78, Dekel86}. The low-mass galaxies observed at the present epoch are therefore best understood either as analogs of those primordial systems or as the surviving ``leftovers'' that have grown slowly and avoided incorporation into more massive structures. Their abundance in the local universe is well established by the steep low-mass end of the stellar mass function \citep{Baldry12}. Despite this prevalence, the physical processes governing their evolution remain less well constrained than for more massive galaxies, largely due to observational challenges in obtaining high-quality spectroscopic data for faint systems. Because of their shallow gravitational potentials, LMGs are particularly sensitive to internal feedback from supernovae and stellar winds, as well as to external processes such as ram pressure stripping and tidal interactions. This sensitivity makes them ideal laboratories for studying baryon cycling, star formation regulation, and the enrichment of the circumgalactic and intergalactic media \citep{Tremonti04, Robertson10, Peeples11, Lin23}.

Star formation in galaxies is commonly parameterized through scaling relations between global properties such as stellar mass ($\rm M_\ast$), SFR, and dust content. In massive galaxies, the star-forming main sequence (SFMS) relation between $\rm M_\ast$ and SFR is well established out to high redshifts, with a slope slightly shallower than unity and modest scatter \citep[e.g.,][]{Noeske07,Whitaker14,Speagle14, Kurczynski16, Bisigello18, Leja22}. At the low-mass end, the situation is less clear: while several studies find that LMGs largely follow an extrapolation of the SFMS with increased apparent scatter \citep[e.g.,][]{Boogaard18}, much of what is known about SFHs at $\rm M_\ast\!\lesssim\!10^9\,M_\odot$ comes from Local Group/Local Volume dwarfs—systems influenced by the Milky Way–Andromeda environment and not fully representative of the field \citep{Tolstoy09, Weisz14}. Moreover, discrepancies between instantaneous (H$\alpha$) and $\sim\!100$\,Myr–averaged (UV/SED) SFR indicators do not uniquely imply discrete ``bursts''; similar offsets can arise from smoothly rising SFHs and differences in tracer timescales and associated systematics, as shown by forward modeling of dwarf-galaxy SFHs \citep{Weisz12} and recognized in early work on irregulars (\cite{Gallagher84}; see also \cite{Lee09,Emami19, Mehta23}). Environmental trends in wide surveys further underscore the need to control for selection when interpreting ``burstiness'' \citep{Geha12}.

Dust attenuation is another key parameter shaping our interpretation of galaxy properties, as it affects measurements of SFRs, stellar populations, and emission-line diagnostics. In massive galaxies, the relationship between stellar and nebular attenuation has been extensively studied, with nebular regions generally found to be more obscured than the stellar continuum \citep{Calzetti00, Garn10, Reddy15, Shivaei20, Chartab24}. This difference is typically interpreted in the framework of a two-component dust model \citep{Charlot00}, wherein young stars in H\,{\sc ii} regions are embedded in dense, dusty birth clouds while the bulk of the stellar light passes through more diffuse ISM dust. Such correlations between nebular-to-stellar reddening and other galaxy properties—such as stellar mass, SFR, sSFR (SFR per unit stellar mass), and gas-phase metallicity—offer valuable insights into dust geometry and the ISM \citep{Garn10, Reddy15, Hemmati15, Shivaei20}. However, these studies predominantly focus on massive galaxies, leaving the relationships between dust content, dust geometry, and other galaxy properties far less explored in low-mass, low-metallicity systems, which are expected to have lower overall dust content, more irregular dust distributions, and potentially different attenuation characteristics.

In this paper, we present a detailed spectroscopic analysis of a statistically significant sample of LMGs at $z \sim 0.15$ in the COSMOS field, based on deep observations with the IMACS spectrograph \citep{Dressler11} on the Magellan telescope. We utilize rest-frame optical emission lines to investigate their physical properties, focusing on two key aspects: the star formation rate–stellar mass (SFR–$\rm M_\ast$) relation and dust attenuation characteristics, including the differential reddening between nebular and stellar components. We further compare the LMGs to a control sample of intermediate-mass galaxies (IMGs; $\rm 10^9 \lesssim M_\ast/M_\odot \lesssim 10^{10}$) at $z \sim 0.35$, observed with an identical spectroscopic setup \citep{Patel23}. 

The paper is structured as follows: Section~\ref{Data} describes the sample selection, data and spectroscopic observations. Section~\ref{analysis} outlines our SED fitting and emission-line measurements. Section~\ref{results} presents the SFR–$M_\ast$ relation, dust trends, and nebular-to-stellar reddening ratios. In Section~\ref{discussion}, we discuss the implications for feedback, star formation regulation, and dust geometry in LMGs, and summarize our conclusions. 

Throughout this paper, we assume a flat $\Lambda$CDM cosmology with $H_0=70 \rm \ kms^{-1} Mpc^{-1}$, $\Omega_{m_{0}}=0.3$, and $\Omega_{\Lambda_{0}}=0.7$. All the physical parameters are measured assuming a \cite{Kroupa01} initial mass function (IMF) and the magnitudes are reported on the AB system \citep{Oke83}.

\begin{figure*}
    \centering
    \subfloat{\includegraphics[width=\textwidth,clip=True, trim=0cm 0cm 0cm 0cm]{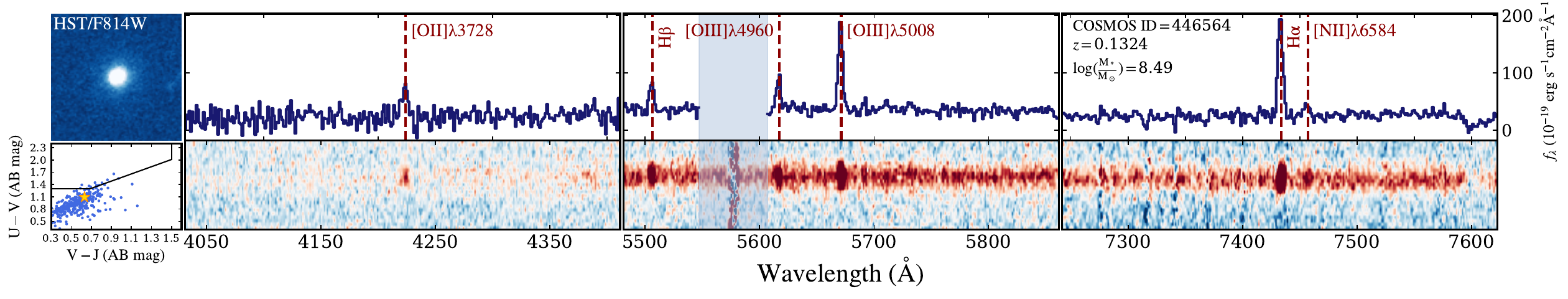}}
    \qquad
    \subfloat{\includegraphics[width=\textwidth,clip=True, trim=0cm 0cm 0cm 0cm]{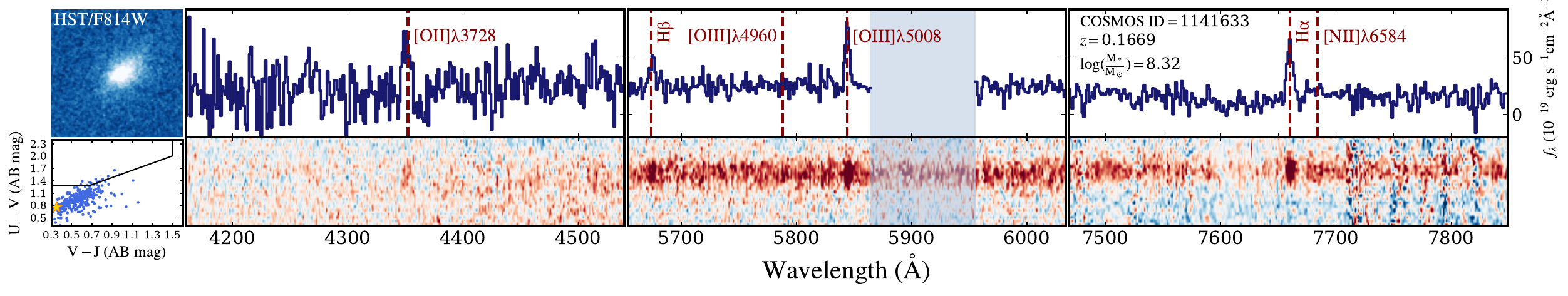}}
    \qquad
    \subfloat{\includegraphics[width=\textwidth,clip=True, trim=0cm 0cm 0cm 0cm]{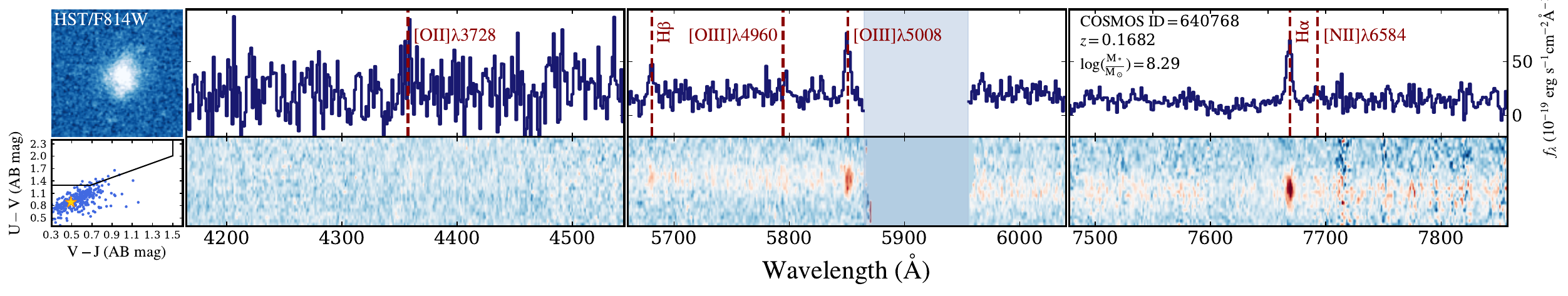}}
    \qquad
    \subfloat{\includegraphics[width=\textwidth,clip=True, trim=0cm 0cm 0cm 0cm]{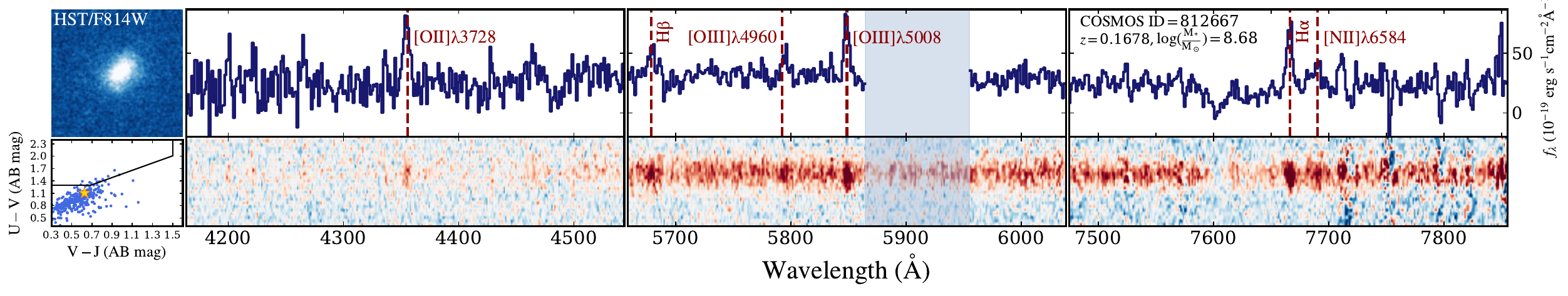}}
    \caption{1D and 2D observed-frame spectra of four example LMGs observed with a $\sim$3.5-hour exposure using IMACS/Magellan. In each row, the UVJ diagram (bottom left) displays the entire sample of LMGs with the example galaxy marked (yellow star). The F814W/HST cutout of the galaxy is also shown in the top left.}\label{examples} 
\end{figure*}

\section{Data}\label{Data}

In this paper, we study a sample of LMGs in the Cosmic Evolution Survey \citep[COSMOS;][]{Scoville07} field. Our galaxies are initially selected based on two criteria on their stellar mass and photometric redshift ($z_{phot}$):\\

1. $10^8\leq \rm M_*/M_\odot \leq 10^9$;

2. $0.1\leq z_{phot} \leq 0.2$.\\

\noindent Spectroscopy of these galaxies was acquired in 2019, 2023, and 2024. We discuss details of the observations in the next section. The galaxies observed in 2019 were selected from \cite{Muzzin13a}, with their $z_{phot}$ calculated using the {\tt EAZY} code \citep{Brammer08} and their stellar masses determined through the SED fitting code of {\tt FAST} \citep{Kriek09}. However, the more recent sample was selected from the {\tt CLASSIC} catalog of COSMOS2020 \citep{Weaver22}. The stellar mass and $z_{phot}$ of these sources are selected from a combination of {\tt EAZY} \citep{Brammer08} and {\tt LePhare} \citep{Arnouts02, Ilbert06} solutions of SED fitting. 

For all the sources initially selected from the two catalogs of \cite{Muzzin13a} and \cite{Weaver22} based on their stellar mass and $z_{phot}$, we then perform SED fitting by fixing their redshifts to the spectroscopic values to recalculate their physical properties (Section \ref{sed-fitting}).

\subsection{IMACS Observations}\label{observations}

Deep optical spectroscopic observation of our galaxies is obtained using the Inamori-Magellan Areal Camera and Spectrograph \citep[IMACS;][]{Dressler11} at the Magellan Baade telescope over the observing programs during 2019, 2023, and 2024. All 12 masks were observed under clear conditions with an average seeing of $\sim 0.6 {\arcsec}$. The typical exposure time for each mask was $\sim 210$ minutes.

$f/2$ spectroscopy with 300 $\rm mm^{-1}$ grism disperser and the the appropriate spectroscopic filter was used in all runs, yielding a spectral resolving power of R $\sim$ 1000 (resolving $\sim$6 \AA{} at 6000 \AA{}). The slit widths were $1 {\arcsec}$, which corresponds to $1.84$–$3.30$~kpc across $0.1\le z\le0.2$ (2.61~kpc at $z=0.15$) and $4.45$–$5.37$~kpc across $0.3\le z\le0.4$ (4.94~kpc at $z=0.35$). The wide range of wavelength coverage (3700-9500\,\AA{}) allows the detection of strong emission lines, such as $\rm H\alpha$ and $\rm H\beta$, simultaneously.

\subsection{Data Reduction}\label{data_reduction}

To reduce the spectroscopic data, we utilize the Carnegie Python Distribution (CarPy\footnote[1]{\url{https://code.obs.carnegiescience.edu/}}), which is a powerful tool tailored for IMACS observations \cite[see e.g.,][]{Kelson00, Kelson03}. The pipeline stacks individual frames, subtracts the sky background, rectifies and extracts 2D spectra. Following this, we apply the optimal extraction algorithm to obtain 1D spectra of galaxies and associated errors \citep[e.g.,][]{Horne86}. Spectroscopic redshifts of galaxies are determined based on strong emission lines in the galaxy spectra, such as H$\alpha$, [O{\sc iii}]$\lambda 5008$, and [O{\sc ii}]$\lambda 3727$. Figure \ref{examples} displays both 2D and 1D spectra for selected LMGs, along with their positions on the UVJ diagram and their F814W/HST postage stamps.

\subsection{Intermediate-Mass Control Sample}\label{control}

As a comparison, a similarly observed sample of IMGs is also added to our LMG sample. The details of the observation and sample selection of these galaxies are discussed extensively in \cite{Patel23}, but we provide a brief summary here. IMGs had been initially selected from \cite{Muzzin13a} based on their $z_{phot}$ and stellar mass, such that $10^9\leq \rm M_*/M_\odot \leq 10^{10}$, and $0.3\leq z_{phot} \leq 0.4$. The stellar masses were the product of {\tt FAST} SED fitting code \citep{Kriek09}, and the photometric redshifts were measured with the {\tt EAZY} code \citep{Brammer08}. Magellan IMACS spectroscopy, along with the measurement of spectroscopic redshifts and emission line fluxes, is detailed in \citet{Patel23}. For a full description, we refer the reader to this work. 

\begin{figure}
\centering
\includegraphics[width=0.45\textwidth,clip=True, trim=0cm 0cm 0cm 0cm]{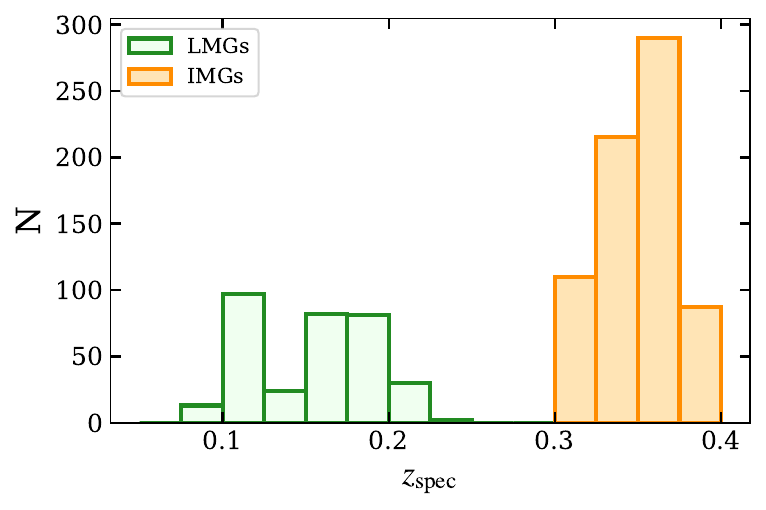}
\caption{Spectroscopic redshift distribution of both LMGs and IMGs.}\label{z_hist} 
\end{figure} 

Figure \ref{z_hist} shows the spectroscopic redshift distribution for both LMG and IMG samples. Our main sample consists of 329 LMGs (green histogram) and 702 IMGs (orange histogram) at $0.1 \lesssim z \lesssim 0.2$ and $0.3 \lesssim z \lesssim 0.4$, respectively.

\section{Analysis}\label{analysis}

\begin{figure*}
    \centering
    \subfloat{{\includegraphics[width=0.47\textwidth,clip=True, trim=0cm 0cm 0cm 0cm]{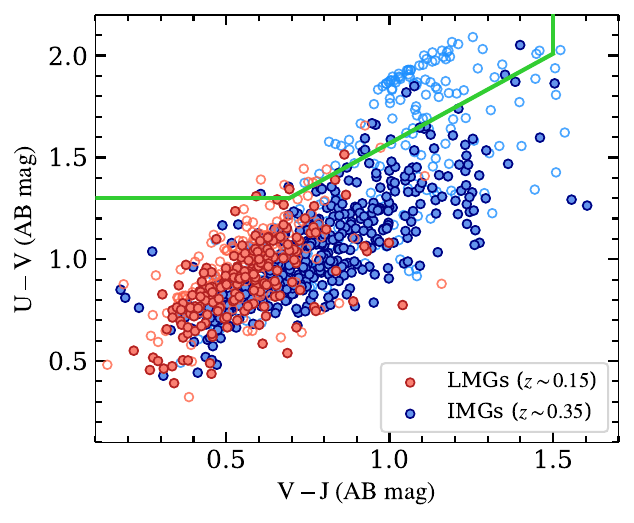} }}%
    \subfloat{{\includegraphics[width=0.51\textwidth,clip=True, trim=0cm 0cm 0cm 0cm]{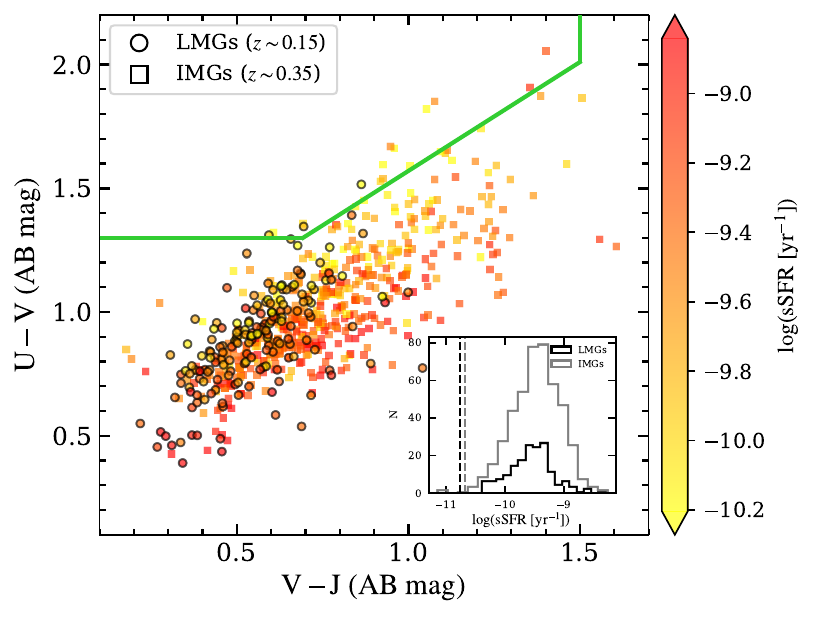} }}%
    \qquad
    \caption{\textit{Left}: The rest-frame UVJ plane derived from SED fitting for both LMGs (red) and IMGs (blue). The filled red (blue) colors correspond to LMGs (IMGs) with $3\sigma$ detection in their H$\alpha$ and H$\beta$ lines. The green line indicates the boundary between quiescent galaxies and star-forming galaxies. \textit{Right}: The same galaxies shown in the left panel, limited to objects with $\geq 3\sigma$ detections in both H$\alpha$ and H$\beta$, color-coded by their sSFR. The inset shows the sSFR distributions of the LMG and IMG samples, with the vertical dashed line indicating the star-forming/quiescent division at $\mathrm{sSFR} = 0.2/t_U(z)$ evaluated at the median redshift of each sample.}\label{uvj} 
\end{figure*}

\subsection{SED Fitting}\label{sed-fitting}

Utilizing the Bayesian Analysis of Galaxies for Physical Inference and Parameter EStimation \citep[{\tt Bagpipes};][]{Carnall18}\footnote[2]{\url{https://bagpipes.readthedocs.io/en/latest/}} framework, we fit the UV to mid-infrared SEDs of both LMGs and IMGs to derive their physical properties, and to model the high resolution spectrum of Balmer absorption underneath the emission lines. The observed photometry for these galaxies is taken from the COSMOS2020 {\tt CLASSIC} catalog \citep{Weaver22}, which is corrected for galactic dust and rescaled to the total flux for the 3{\arcsec} diameter aperture photometry. The SED fitting uses broad-band photometry spanning GALEX NUV; CFHT {\it u}; Subaru/HSC {\it g, r, i, z, y}; UltraVISTA {\it Y, J, H, Ks}; and Spitzer/IRAC channels 1 and 2 (3.6 and 4.5~$\mu$m).

We adopt photometry from the COSMOS2020 {\tt CLASSIC} catalog rather than the {\tt FARMER} catalog. For the optical magnitude range of our sample (HSC $i \sim$ 20--23~AB), we find that {\tt CLASSIC} and {\tt FARMER} broad-band fluxes are highly consistent for all photometric bands used in the SED fitting. As a result, the choice of photometric catalog does not significantly affect the results/conclusions of our analysis. In addition, {\tt CLASSIC} aperture photometry is particularly robust for LMGs with irregular, clumpy, or asymmetric morphologies, which are common among our sample.

{\tt Bagpipes} employs the 2016 version of the \cite{Bruzual03} stellar population synthesis (SPS) models. Additionally, the nebular emission modeling within the code utilizes the 2017 version of the {\tt Cloudy} photoionization code \citep{Ferland17} based on the \cite{Byler17} methodology. In the process of SED fitting, we fix the redshifts to their spectroscopic values and adopt a delayed exponentially declining SFH of the form $\rm t\,e^{-t/\tau}$. We impose a uniform prior on the star formation e-folding time-scale ($\tau$), ranging from 0.3 to 10 Gyr. For dust attenuation, we adopt the \citet{Calzetti00} attenuation curve with $A_V \in (0, 2)$ for the IMG sample. For LMGs, we instead employ the Small Magellanic Cloud (SMC) extinction curve \citep{Gordon03}, motivated by observational and theoretical studies showing that SMC-like dust laws provide a better description of low-metallicity, low-mass systems \citep[e.g.,][]{Salmon16, Reddy18,Salim18,Shivaei20}. Additionally, we allow the stellar metallicity to vary over the range $0 < \rm Z/Z_\odot < 2.5$, assuming a logarithmic prior.

The best-fit SED model for each galaxy, along with stellar masses and other physical parameters (e.g., SFR and stellar dust attenuation), is obtained from SED fitting. The rest-frame UVJ colors are also measured for both LMGs and IMGs. In the left panel of Figure~\ref{uvj}, we demonstrate the position of our galaxies on the rest-frame UVJ plane. The green line in the figure separates the population of quiescent galaxies from the star-forming ones based on the definition from \cite{Muzzin13b}. The filled red (blue) data points correspond to LMGs (IMGs) with $3\sigma$ detections in their H$\alpha$ and H$\beta$ lines used in the present work. 

As shown in the figure, the vast majority of our LMGs and IMGs lie in the star-forming region of the UVJ plane, with only a small minority ($\sim 1\%$ of LMGs and $\sim 2\%$ of IMGs) classified as quiescent. We exclude these quiescent galaxies from both samples. Adopting an sSFR-based criterion \citep{Pacifici16}, ${\rm sSFR}>0.2/t_U(z)$ (where ${\mathrm{sSFR}} \equiv {\mathrm{SFR}}_{\mathrm{H}\alpha}/\rm {M}_*^{\mathrm{SED}}$ and $t_U(z)$ is the age of the Universe at redshift $z$), yields a relatively consistent classification, with disagreements limited to a few objects ($\lesssim2\%$) in the total sample, as illustrated by the sSFR distribution in the right panel of Figure~\ref{uvj}.

We note that UVJ-based classifications are known to be imperfect, with potential contamination from dusty star-forming galaxies near the quiescent boundary \citep{Leja19c}. However, in the present analysis, this effect is minimal given the consistency between UVJ- and sSFR-based classifications.


\subsection{Emission Line Measurements}\label{elm}

Given the wide wavelength coverage of IMACS, we measure the emission line fluxes in three wavelength ranges with strong emission lines:\\

1. [O{\sc ii}]$\lambda \lambda 3727, 3730$;

2. H$\beta$, [O{\sc iii}]$\lambda \lambda 4960, 5008$;

3. H$\alpha$, [N{\sc ii}] $\lambda \lambda 6550, 6585$.\\

\noindent In each wavelength range, we simultaneously fit multiple Gaussians to the 1D spectra of galaxies to extract the desired emission line fluxes. The fluxes of H$\alpha$ and H$\beta$ emission lines are corrected for underlying Balmer absorption using the best-fit SED models obtained in Section \ref{sed-fitting}. We measure the line fluxes by integrating the best-fit Gaussian function over the 1D spectrum. Uncertainties are taken from the covariance matrix of the multi-Gaussian fit, with continuum and per-pixel variance propagated to the integrated line flux.


\section{Results}\label{results}

\subsection{SFR-$\rm M_*$ Relation}\label{sfr_m_sec}

Out of 329 (702) LMGs (IMGs) at $z \sim 0.15$ ($z \sim 0.35$), 174 (448) of them have signal-to-noise (S/N) $\geq 3$ detection in their H$\alpha$ and H$\beta$ emission lines (these galaxies are shown with bold colors on the UVJ diagram in the left panel of Figure \ref{uvj}). We derive the H$\alpha$-based SFRs (SFR$_{\text{H}\alpha}$) for our LMG and IMG samples using the \cite{Kennicutt98} calibration for a \citet{Kroupa01} IMF: $\rm SFR_{\text{H}\alpha}(M_\odot\,yr^{-1})=5.5 \times 10^{-42}\ L_{H \alpha}(erg/s)$. H$\alpha$ fluxes used for the SFR measurements are corrected for attenuation using the Balmer decrement (see Section \ref{dust_sec}). Uncertainties on SFR$_{\mathrm{H}\alpha}$ include both line–flux measurement errors and propagated $E(B\!-\!V)_{\rm neb}$ uncertainties.

\begin{figure}
\centering
\includegraphics[width=0.44\textwidth,clip=True, trim=0cm 0cm 0cm 0cm]{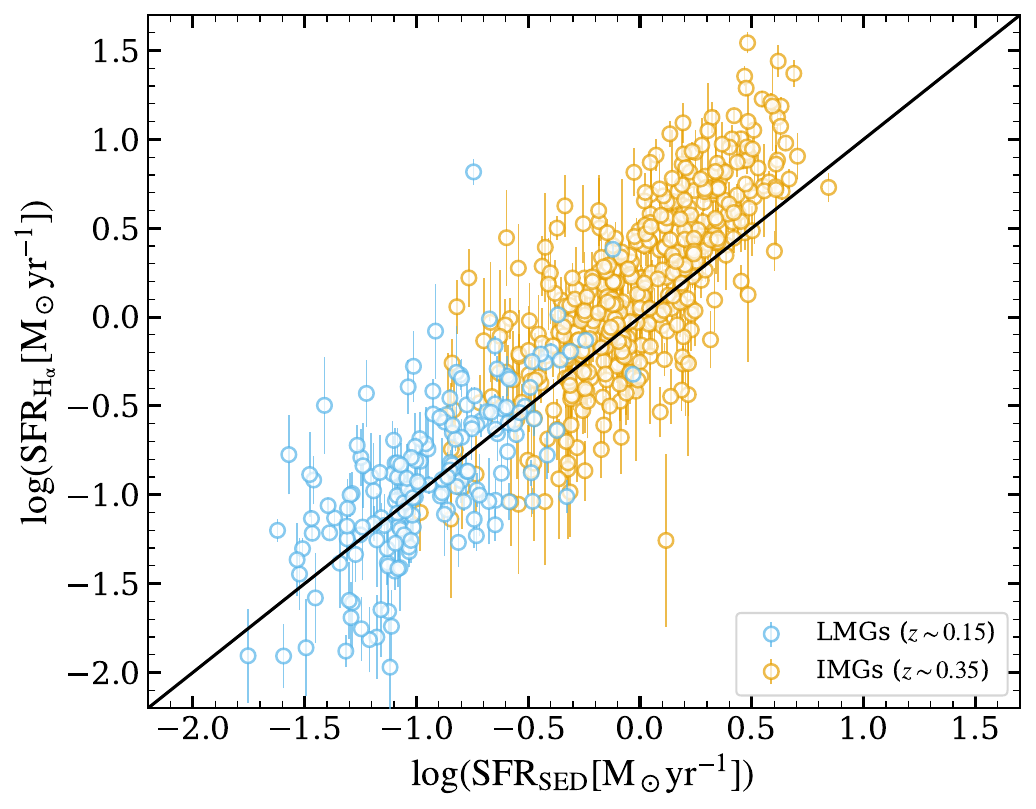}
\caption{Comparison between the SFRs estimated from SED fitting ($\rm SFR_{SED}$) with those based on the H$\alpha$ luminosity (SFR$_{\text{H}\alpha}$) for LMGs (blue circles) and IMGs (orange circles). The black line represents a unit slope, indicating equality between SFR$_{\text{H}\alpha}$ and $\rm SFR_{SED}$.}\label{sfr_comp} 
\end{figure}

\begin{figure*}
\centering
\includegraphics[width=0.81\textwidth,clip=True, trim=0cm 0cm 0cm 0cm]{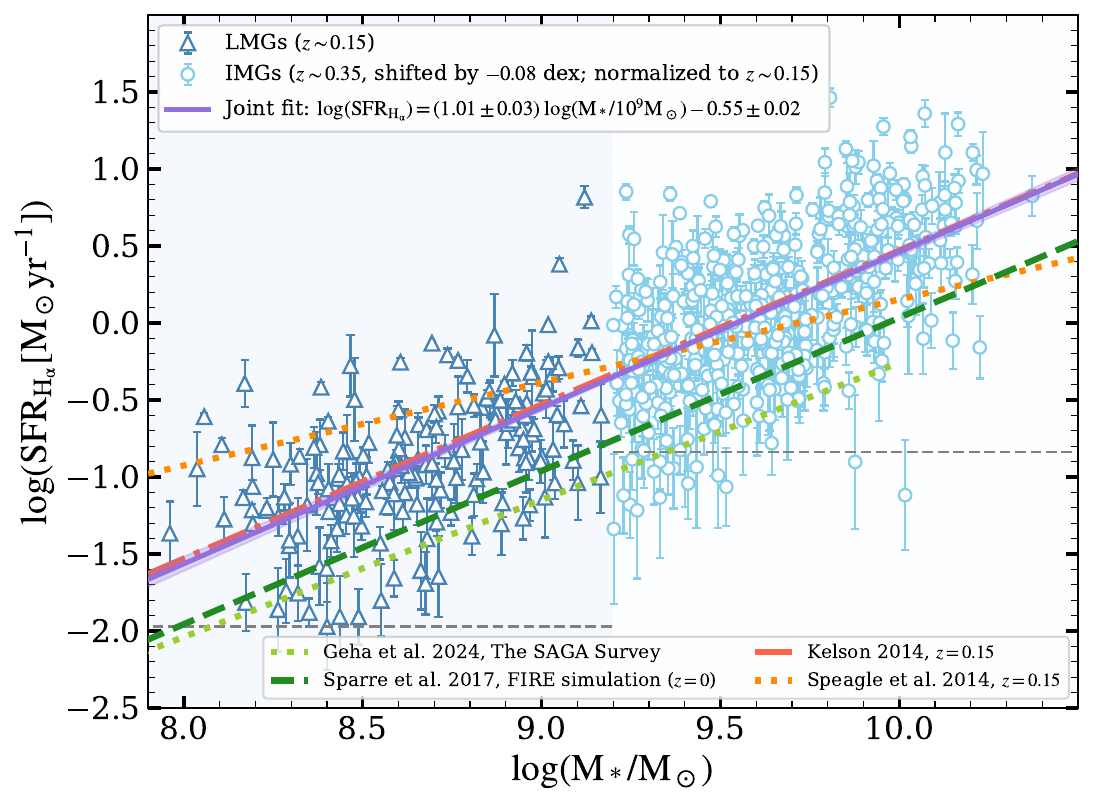}
\caption{SFR$_{\text{H}\alpha}$-$\rm M_\ast$ relation for LMGs and IMGs. Blue triangles show LMGs ($z\sim 0.15$) and light–blue circles show IMGs shifted by $-0.08$\,dex in $\log\mathrm{(SFR)}$ to the $z\sim 0.15$ frame. The solid purple line is the joint fit (ordinary least squares on $\log\mathrm{SFR}_{\mathrm{H}\alpha}$ vs.\ $\rm \log M_\ast$), yielding $a=1.01\pm0.03$ and intercept ${-0.55\pm0.02}$; the shaded envelope indicates the 68\% confidence interval from 1000 bootstrap realizations. For context we overlay other relations from literature: \citet{Speagle14} at $z=0.15$ (orange dotted), the constant-sSFR track from \citet{Kelson14} at $z=0.15$ (red dash–dot), the SAGA dwarfs line from \citet{Geha24} (green dotted), and the FIRE zoom-in locus from \citet{Sparre17} at $z=0$ (dark green dashed). Gray dashed horizontal segments indicate the representative S/N\,$\geq$\,3 detection floors; the IMG threshold is shown with the same $-0.08$\,dex shift for consistency. Error bars on points denote $1\sigma$ uncertainties in $\log\mathrm{(SFR}_{\mathrm{H}\alpha})$.}\label{sfr_m} 
\end{figure*}

\begin{figure*}
\centering
\includegraphics[width=0.95\textwidth,clip=True, trim=0cm 0cm 0cm 0cm]{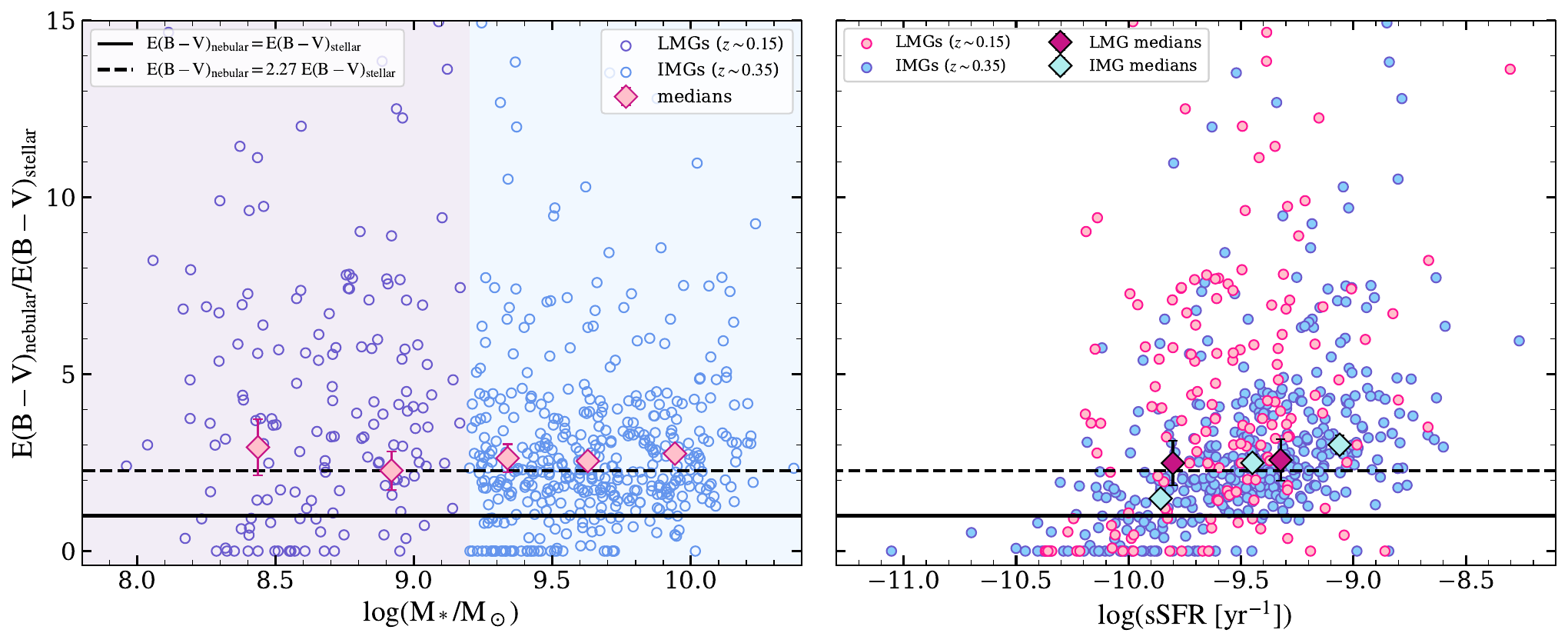}
\caption{Left: Variation of nebular to stellar color excess with stellar mass for LMGs (purple circles) and IMGs (blue circles). Median values for each sample are represented in two stellar mass bins using pink diamonds. Error bars indicate the uncertainty of these medians. Right: Nebular to stellar color excess as a function of sSFR for LMGs (pink circles) and IMGs (blue circles). Median values, along with their uncertainties, are depicted with magenta (blue) diamonds for the LMG (IMG) sample in two bins of sSFR. The solid and dashed lines in both panels show identical nebular and stellar color excess, and the relation between the two reddening from \cite{Calzetti00} ($E(B-V)_{\rm nebular}=2.27\ E(B-V)_{\rm stellar}$), respectively.}\label{ebv_mass_ssfr} 
\end{figure*}

We also estimate SFRs from SED fitting ($\rm SFR_{SED}$). While the SFR coming from H$\alpha$ flux measurements traces the instantaneous star formation activity within galaxies, the SED-derived SFR is the integrated SFR of the galaxy over 100 Myrs and is more susceptible to degeneracies present in SED modeling. Nevertheless, as shown in Figure \ref{sfr_comp}, both indicators are in reasonable agreement, with a normalized median absolute deviation ($\rm \sigma_{NMAD}$) of $\sim 0.3$ for each sample, consistent with the results of \citet{Patel23} for IMGs. As an external benchmark on short–timescale variability for IMGs, \cite{Patel23} compared dust-corrected H$\alpha$ and rest-frame 2800\,\AA{} UV luminosities and found a very small intrinsic scatter between their offsets from the respective sequences ($\sigma_\delta \lesssim 0.03$\,dex), which limits recent SFR variability to $\lesssim 2$ over 200\,Myr and $\lesssim 30\%$ over 20\,Myr—disfavoring strongly bursty SFHs and favoring an ensemble of gently varying histories. This behavior is naturally connected to the dependence of nebular attenuation on sSFR, which we examine explicitly in Section \ref{dust_sec} and Figure \ref{ebv_mass_ssfr}.

Figure \ref{sfr_m} presents the relation between SFR$_{\text{H}\alpha}$ and stellar mass for LMGs and IMGs. Following \citet{Kelson14}, we assume that the median sSFR declines as the inverse of cosmic time, $\langle{\rm sSFR}\rangle \propto 1/t_U(z)$, which implies a mass–independent evolution in the normalization of the SFMS. Under our assumed cosmology, this corresponds to a $\sim0.08$ dex decrease in $\log{\rm (SFR)}$ from $z\!\sim\!0.35$ to $z\!\sim\!0.15$, so we shift the higher–$z$ sample (IMGs) by $-0.08$ dex to place both samples on a common ($z\!\sim\!0.15$) evolutionary frame. 

\begin{figure*}
\centering
\includegraphics[width=0.81\textwidth,clip=True, trim=0cm 0cm 0cm 0cm]{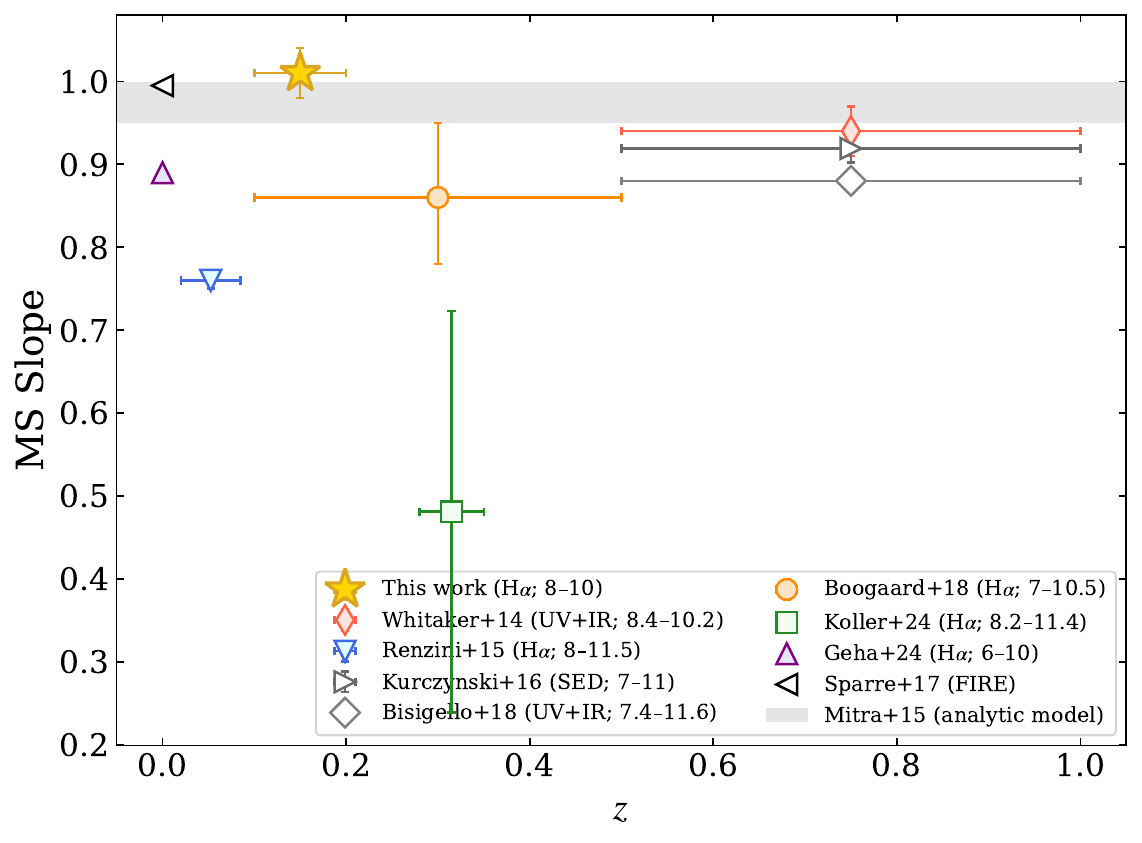}
\caption{Slope of the SFMS as a function of redshift, where the MS slope $a$ is defined by $\log{\rm (SFR)}=a\,\rm \log (M_\ast/10^9 M_\odot) +$$b$. The gold star marks this work (joint fit to LMGs and IMGs), yielding $a=1.01\pm0.03$ at $z\sim 0.15$. Literature points (symbols/labels) include \citet{Whitaker14} (UV+IR), \citet{Renzini15} (H$\alpha$), \citet{Kurczynski16} (SED), \citet{Bisigello18} (UV+IR), \citet{Boogaard18} (H$\alpha$), \citet{Koller24} (H$\alpha$), \citet{Geha24} (H$\alpha$), and the FIRE zoom-ins from \citet{Sparre17}. Horizontal bars indicate each study’s redshift range; vertical bars show the reported $1\sigma$ uncertainty on $a$. The gray band denotes the analytic expectation for the SFMS (i.e., $a\sim1$) from \citet{Mitra15}. Numbers in the legend denote the stellar-mass interval (in $\rm \log(M_\ast/M_\odot)$) over which each slope was measured.}
\label{sfr_m_slope}
\end{figure*}

We model the SFMS as a linear relation in log space,
$\log(\mathrm{SFR}_{\mathrm{H}\alpha}) = a\,\log \rm (M_{\ast}/10^9 M_\odot)+$$b$. Parameters $(a,b)$ are obtained by ordinary least squares (minimizing vertical residuals) using \texttt{curve\_fit} \citep{scipy}. To obtain robust uncertainties on $(a,b)$, we employ bootstrap resampling; we draw 1000 samples with replacement from the combined (LMGs + shifted IMGs) dataset, refit the linear relation to each realization, and report the median slope and intercept along with their 16th–84th percentile ranges. The shaded band in Figure \ref{sfr_m} shows the 68\% confidence region derived from the distribution of bootstrap regression lines. The regression uses only objects with S/N\,$\geq$\,3 in both H$\alpha$ and H$\beta$; we do not incorporate the S/N selection cuts into the fit. Instead, we indicate ``representative'' S/N\,$\geq$\,3 detection floors in the SFR--$\rm M_\ast$ plane (horizontal gray dashed lines in each mass range), computed from the median 1$\sigma$ line–flux depths and median nebular $A_V$ for each sample (and shifted by $-0.08$\,dex for IMGs to match the joint–fit frame).

A single linear fit to the combined sample (purple line) yields a slope of $1.01\pm0.03$ and intercept ${-0.55\pm0.02}$ in the $z\sim 0.15$ frame, i.e., SFR increases systematically with stellar mass in all mass ranges, following the expected SFMS trend \citep[e.g.,][]{Speagle14}. We estimate the intrinsic scatter of $\sigma_{\rm int}\!\sim\!0.4$\,dex, which is identical to the expectation for an $\rm H=1$ stochastic process and scale-free gravitational collapse \citep{Kelson14, Kelson20}. 

For comparison, Figure \ref{sfr_m} also overlays other relations from literature as reference. We plot the SAGA dwarfs best–fit line from \citet{Geha24} (light–green dotted), which anchors the low–mass, low–$z$ regime, and the constant–sSFR guide from \citet{Kelson14} evaluated at $z=0.15$ (red dot–dashed). We further show the time–evolving ``consensus'' main sequence from \citet{Speagle14} evaluated at our sample redshift (orange dotted). As a theoretical check, we include the FIRE zoom–in locus at $z=0$ from \citet{Sparre17} (green dashed). In FIRE, the ensemble FUV–based main sequence has an approximately unity slope over our mass range, but the H$\alpha$–based SFRs on $\sim$10\,Myr timescales exhibit larger variability, especially below $\rm M_\ast\!\sim\!10^{10}\,M_\odot$, indicative of bursty SFHs. By contrast, our observed intrinsic scatter of $\sigma_{\rm int}\!\sim\!0.4$\,dex is comparable in both mass ranges and aligns with previous observational constraints for massive galaxies \citep[e.g.,][]{Kurczynski16}. Thus, if one infers SFH ``burstiness'' from the SFMS scatter alone, our LMGs exhibit variations similar to those of more massive star–forming systems.

We emphasize that the SED fitting in this work is not used to infer SFHs. The delayed exponentially declining SFH adopted in the SED fitting is chosen to provide robust stellar masses, stellar continuum attenuation, and Balmer absorption corrections, rather than to recover time-resolved SFHs. Previous studies have shown that stellar mass estimates are relatively insensitive to SED fitting assumptions \citep{Pacifici23}. As a result, our discussion of bursty versus smoothly varying star formation does not rely on the SED-based SFH itself. Instead, constraints on burstiness are based on spectroscopic diagnostics, including the scatter in the SFR--$\rm M_*$ relation (SFMS) derived from H$\alpha$-based SFRs.

\begin{figure*}
\centering
\includegraphics[width=\textwidth,clip=True, trim=0cm 0cm 0cm 0cm]{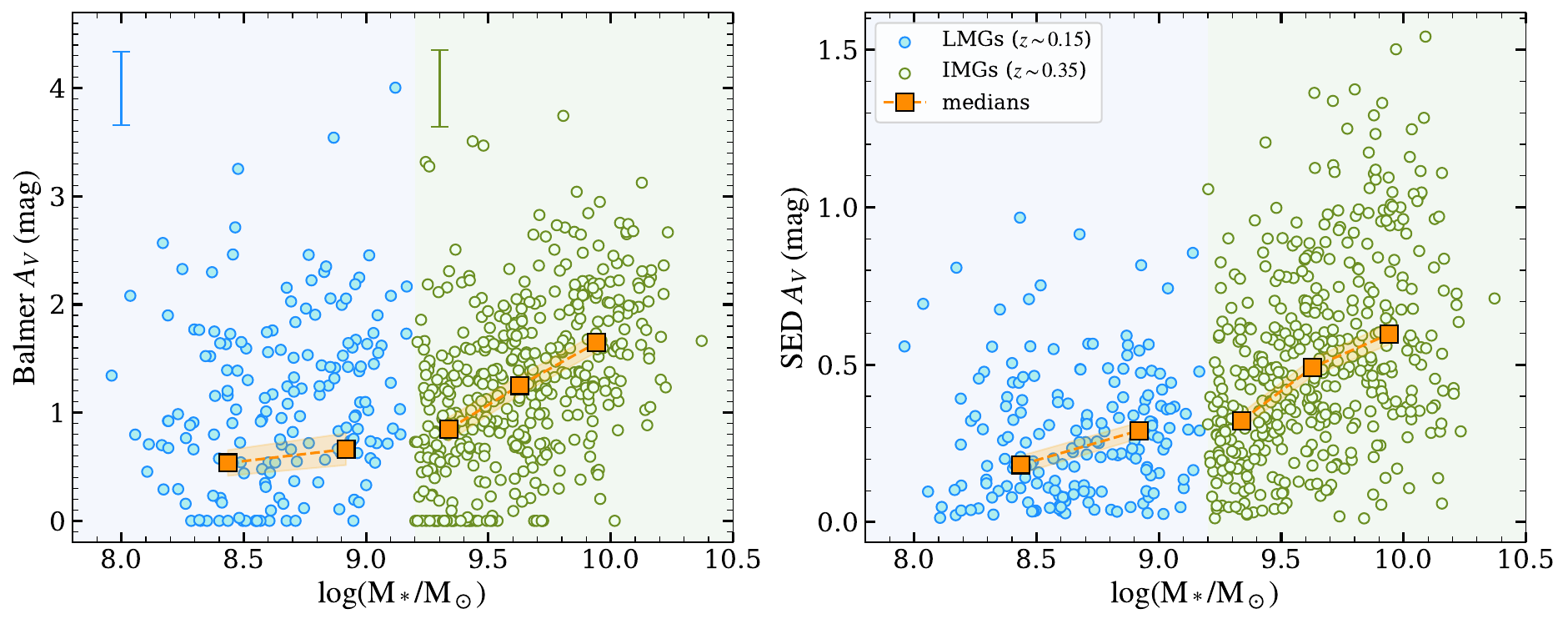}
\caption{Left: Balmer $A_V$ plotted against stellar mass for LMGs (blue circles) and IMGs (green circles). Orange squares indicate the median trend within each sample. The typical error bar of each sample is displayed in the top-left corner. Right: SED-derived $A_V$ versus stellar mass for LMGs (blue circles) and IMGs (green circles), with the median trend shown as orange squares across different stellar mass bins. The shaded regions around the median values in each panel indicate the uncertainty associated with these medians.}\label{AV_M} 
\end{figure*}

Finally, Figure \ref{sfr_m_slope} compiles the comparison measurements of the SFMS slope as a function of redshift from the literature; the legend indicates the SFR indicator adopted along with the stellar mass range in each study. At very low redshift, we include the SDSS analysis of \citet{Renzini15}, who define the SFMS as the ridge line of the three–dimensional (SFR, $\rm M_\ast$, number) distribution over $0.02<z<0.085$. At $0.1<z<0.5$, we show the VLT/MUSE results of \citet{Boogaard18}, who derive dust–corrected H$\alpha$–based SFRs via the Balmer decrement for LMGs and fit the MS slope across $\sim10^{7}$–$10^{10.5}\,\rm M_\odot$. For $0.5<z<1$, we plot determinations from \citet{Whitaker14} (based on a mass–complete 3D–HST/CANDELS sample with UV+IR SFRs), \citet{Kurczynski16} (hierarchical fits to SED–based SFRs allowing for intrinsic scatter), and \citet{Bisigello18} (UV with mid/far–IR constraints in CANDELS/GOODS–S). We also show MAGPI IFS results from \citet{Koller24} at $0.28<z<0.35$, noting that their global (integrated) MS slope is less tightly constrained than their resolved rSFMS measurements. As additional anchors, we include the FIRE hydrodynamical prediction at $z\sim 0$ \citep{Sparre17} and the SAGA survey result for nearby satellites \citep{Geha24}. The gray band denotes the slope range from the analytic equilibrium model of \citet{Mitra15}.

Our measured MS slope is consistent with the $\sim$unity value in the FIRE simulations \citep{Sparre17} and with the analytic expectations of \citet{Mitra15}; it also agrees with the \citet{Whitaker14} slope value despite that study’s higher redshift. A slope consistent with unity in the SFMS implies an approximately mass-invariant sSFR over the fitted mass range ($\mathrm{SFR}\propto \rm M_\star$). This indicates that galaxies in this stellar mass range grow their stellar content in a self-similar manner, forming stars at rates directly proportional to their existing stellar mass.

We note that several literature studies shown extend to stellar masses well above the range probed here, where a turnover or flattening of the SFMS is often observed; slopes reported over broader mass ranges may therefore differ systematically from those measured when restricting the fit to lower stellar masses.

\subsection{Dust Attenuation}\label{dust_sec}

We estimate the nebular dust attenuation for both LMGs and IMGs using the Balmer decrement, defined as the H$\alpha$/H$\beta$ line flux ratio. An intrinsic ratio of 2.86 is adopted, corresponding to case B recombination under typical conditions of $T = 10^4\ \rm K$ and $n_e = 10^2\ \rm cm^{-3}$ \citep{Osterbrock06}. To convert the color excess to attenuation ($A_\lambda = k_\lambda E(B\!-\!V)$), we apply the \citet{Calzetti00} attenuation curve for IMGs, while for LMGs we adopt the SMC extinction curve from \citet{Gordon03}, in line with their lower metallicities and the evidence that SMC-like dust laws better characterize such systems \citep[e.g.,][]{Salim18, Shivaei20b}. Uncertainties in attenuation are computed by propagating the flux errors on H$\alpha$ and H$\beta$ using standard error propagation techniques. When the measured Balmer decrement is consistent with or below the intrinsic Case~B value within uncertainties, the inferred $E(B\!-\!V)_{\rm nebular}$ would be formally negative; in such cases we set $E(B\!-\!V)_{\rm nebular}=0$ (and hence $A_V=0$), as negative attenuations are unphysical.

\subsubsection{Trends in Nebular and Stellar Attenuation}

The left panel of Figure~\ref{AV_M} shows Balmer $A_V$ as a function of stellar mass for LMGs and IMGs, with medians in stellar mass bins indicated by orange squares. To compute these medians, we use an inverse–variance–weighted method \citep{Edgeworth1888} that accounts for non-uniform measurement uncertainties in the Balmer decrement. Specifically, within each mass bin we sort the $A_V$ values, accumulate their normalized weights, and define the median as the value at which the cumulative weight first reaches 0.5. The uncertainty on each median is estimated from the weighted scatter of the points in the bin, adjusted by the effective sample size.

IMGs exhibit a clear increase in Balmer $A_V$ with stellar mass, consistent with previous studies of the mass–attenuation relation at both low and high redshifts \citep[e.g.,][]{Garn10,Cullen18}. In contrast, LMGs show systematically lower $A_V$ values and a markedly weaker dependence on stellar mass. This aligns with the known dust deficiency of low-mass, low-metallicity galaxies \citep[e.g.,][]{RemyRuyer14,DeVis19}, which exhibit low dust-to-gas ratios, steeper attenuation curves, and reduced overall attenuation \citep{Reddy18,Shivaei20,Salim20}. The mass-independence of $A_V$ in LMGs likely reflects fundamental limits on dust production and retention in the low-metallicity, feedback-dominated environments of dwarf galaxies.  

A forthcoming analysis of gas-phase metallicities in these LMGs will provide direct constraints on their chemical enrichment and enable a quantitative link among metallicity, dust content, and attenuation; in parallel, morphological measurements from HST/JWST imaging will place these relations in structural context, providing a better understanding of the predominance of local dust around the H\textsc{ii} regions versus the broad interstellar dust distributions within these galaxies.

The right panel of Figure~\ref{AV_M} shows the stellar continuum attenuation (SED $A_V$) from our SED fitting as a function of stellar mass, with median values in stellar mass bins again shown by orange squares. In both LMGs and IMGs, SED $A_V$ increases with stellar mass. However, at fixed stellar mass, SED $A_V$ is systematically lower than the nebular $A_V$, in agreement with previous findings that nebular regions experience higher dust attenuation than the stellar continuum \citep[e.g.,][]{Calzetti00, Garn10, Hemmati15, Qin19, Shivaei20, Patel23, Chartab24}. For the IMG sample, these analyses are also reported in \citet{Patel23}, and our measurements are consistent with theirs.

\subsubsection{Nebular-to-Stellar Reddening Ratios}

To further investigate the differential attenuation in our samples, we plot $E(B-V)_{\mathrm{nebular}}/E(B-V)_{\mathrm{stellar}}$ as a function of stellar mass in the left panel of Figure \ref{ebv_mass_ssfr}. Median values in stellar mass bins for each sample are shown as pink diamonds. The solid horizontal line corresponds to equal reddening of nebular and stellar components, while the dashed line marks the canonical $E(B-V)_{\mathrm{nebular}} = 2.27\,E(B-V)_{\mathrm{stellar}}$ relation for local starbursts \citep{Calzetti00}. 

For both samples the median ratios lie close to this canonical value, in agreement with previous studies of galaxies at similar masses and redshifts \citep[e.g.,][]{Kashino13,Price14,Shivaei20}. Also, there is no statistically significant correlation between the reddening ratio and stellar mass within either population, consistent with earlier findings for more massive systems \citep[e.g.,][]{Hemmati15,Reddy15,Shivaei20}.


We also examine the dependence of the nebular-to-stellar color excess ratio on sSFR in the right panel of Figure \ref{ebv_mass_ssfr}. The sSFR is derived from the dust-corrected H$\alpha$ SFR (SFR$_{\mathrm{H}\alpha}$), as described in Section \ref{sfr_m_sec}. The solid and dashed horizontal lines indicate equal reddening for the nebular and stellar components and the \citet{Calzetti00} relation ($E(B-V)_{\mathrm{neb}} = 2.27\,E(B-V)_{\mathrm{star}}$), respectively. Median values in bins of sSFR are shown as magenta (LMGs) and blue (IMGs) diamonds, with error bars indicating the uncertainty on the median.

At low sSFR ($\sim 10^{-10}\ \rm yr^{-1}$), LMGs exhibit a relatively elevated nebular-to-stellar attenuation ratio compared to IMGs, but the median value remains flat as sSFR increases. This flatness contrasts with the clear positive correlation observed in IMGs. A possible interpretation is that, in LMGs, short birth-cloud clearing times at high sSFR prevents long-lived obscuration \citep[e.g.,][]{Chevance20}. By contrast, the deeper potential wells and higher dust masses of IMGs allow H\,\textsc{ii} regions to remain optically thick as sSFR increases, producing a strong sSFR-dependent rise in the nebular-to-stellar ratio. This possible rise at higher sSFR is qualitatively consistent with results for high-redshift star-forming galaxies \citep[e.g.,][]{Reddy15, Shivaei20}, and may reflect dust geometries in which the youngest, most strongly star-forming regions are more heavily embedded in dust than the older stellar populations.

This sSFR-dependent increase in nebular attenuation provides a natural physical explanation for the offset between $\mathrm{SFR}_{\mathrm{H}\alpha}$ and $\mathrm{SFR}_{\mathrm{SED}}$ observed at high SFR in Figure \ref{sfr_comp}. As sSFR increases, particularly for IMGs, star-forming regions become increasingly embedded and optically thick, enhancing the attenuation of nebular emission relative to the stellar continuum. We emphasize that this behavior reflects changes in dust geometry and the embedding of H\,\textsc{II} regions at fixed star formation activity, rather than strongly bursty SFHs, which are independently constrained to be relatively smooth on short timescales for IMGs \citep{Patel23}.

\section{Discussion \& Summary}\label{discussion}

In this paper, we have presented a detailed spectroscopic analysis of low-mass galaxies (LMGs; $\rm M_\ast \lesssim 10^9\,M_\odot$) at $z\sim 0.15$ in the COSMOS field, based on deep IMACS/Magellan observations, and compared them to a control sample of intermediate-mass galaxies (IMGs; $10^9 \lesssim \rm M_\ast/M_\odot \lesssim 10^{10}$) at $z\sim 0.35$ observed with the same instrumental setup. The final sample includes 174 LMGs and 448 IMGs with $\geq 3\sigma$ detections in both H$\alpha$ and H$\beta$. Using photometry from the COSMOS2020 {\tt CLASSIC} catalog and fixing redshifts to their spectroscopic values, we performed SED fitting with {\tt Bagpipes} to derive stellar masses, stellar dust attenuation, and other physical parameters. Our main findings are summarized below.

\begin{itemize}

\item We measured H$\alpha$-based SFRs (dust-corrected via the Balmer decrement) and compared them to SED-based SFRs, finding good agreement ($\rm \sigma_{NMAD}\sim 0.3$). In the SFR–$\rm M_\ast$ plane, we fit a single linear relation to LMGs and IMGs. The joint ordinary-least-squares fit yields a slope of $a=1.01\pm0.03$ and intercept ${b=-0.55\pm0.02}$, consistent with an approximately linear SFMS at these masses and redshifts. The inferred intrinsic scatter is $\sigma_{\rm int}\sim0.4$\,dex across the full mass range. Both the near-unity SFMS slope and the modest intrinsic scatter point to gradual, scale-free stellar-mass assembly, with no evidence that LMGs are systematically more bursty than IMGs. The comparable scatter in both samples indicates that low-mass systems build up their stellar mass in much the same way as their intermediate-mass counterparts; this level is consistent with the ensemble of generally rising, scale-free SFHs discussed by \cite{Kelson14} and consistent with nonlinear gravitational collapse \citep{Kelson20}.

\item We compared nebular $A_V$ (from the Balmer decrement) and stellar continuum $A_V$ (from SED fitting) for both samples. IMGs show a clear increase in $A_V$ with stellar mass for both nebular and stellar components, while LMGs exhibit lower overall attenuation and a weaker dependence on $M_*$. This difference is consistent with the lower dust content expected in low-metallicity, low-mass galaxies \citep[e.g.,][]{Salim20}. In all cases, stellar $A_V$ is lower than nebular $A_V$, in line with the two-component dust model.

\item The ratio $E(B-V)_{\mathrm{nebular}}/E(B-V)_{\mathrm{stellar}}$ for both LMGs and IMGs is close to the canonical factor of 2.27 from \citet{Calzetti00}. We find no significant correlation of this ratio with stellar mass in either sample, consistent with previous results for more massive galaxies \citep[e.g.,][]{Reddy15,Shivaei20}. When examined as a function of sSFR, the ratio shows a mild tendency to increase toward higher sSFR for IMGs. This trend is qualitatively consistent with scenarios in which dust is preferentially concentrated in the youngest star-forming regions, leading to greater obscuration of nebular emission relative to the stellar continuum \citep[e.g.,][]{Reddy15}. For LMGs, by contrast, the median ratio remains approximately constant, suggesting that the stellar feedback disperses birth clouds efficiently in shallow potentials so that the nebular-to-stellar reddening ratio is roughly sSFR-invariant.

\end{itemize}

Future work will build on this analysis by incorporating gas-phase metallicity measurements and high-resolution morphological information to place the observed dust attenuation trends in a broader physical context, enabling a more direct connection between galaxy morphology, chemical enrichment, and star formation properties in low-mass systems.


\section*{Acknowledgement}

We thank the anonymous referee for providing insightful comments and suggestions that improved the quality of this work. This paper includes data gathered with the 6.5-meter Magellan Telescopes at Las Campanas Observatory, Chile. We thank the staff for their dedication and support.

\bibliography{lmg}

@ARTICLE{Patel23,
       author = {{Patel}, Shannon G. and {Kelson}, Daniel D. and {Abramson}, Louis E. and {Sattari}, Zahra and {Lorenz}, Brian},
        title = "{Constraints on Fluctuating Star Formation Rates for Intermediate-mass Galaxies with H{\ensuremath{\alpha}} and UV Luminosities}",
      journal = {\apj},
         year = 2023,
        month = mar,
       volume = {945},
       number = {2},
          eid = {93},
        pages = {93},
          doi = {10.3847/1538-4357/acb938},
archivePrefix = {arXiv},
       eprint = {2303.04165},
 primaryClass = {astro-ph.GA},
       adsurl = {https://ui.adsabs.harvard.edu/abs/2023ApJ...945...93P}
}

@ARTICLE{Dressler11,
       author = {{Dressler}, Alan and {Bigelow}, Bruce and {Hare}, Tyson and {Sutin}, Brian and {Thompson}, Ian and {Burley}, Greg and {Epps}, Harland and {Oemler}, Augustus, Jr. and {Bagish}, Alan and {Birk}, Christoph and {Clardy}, Ken and {Gunnels}, Steve and {Kelson}, Daniel and {Shectman}, Stephen and {Osip}, David},
        title = "{IMACS: The Inamori-Magellan Areal Camera and Spectrograph on Magellan-Baade}",
      journal = {\pasp},
         year = 2011,
        month = mar,
       volume = {123},
       number = {901},
        pages = {288},
          doi = {10.1086/658908},
       adsurl = {https://ui.adsabs.harvard.edu/abs/2011PASP..123..288D}
}

@ARTICLE{Weaver22,
       author = {{Weaver}, J.~R. and {Kauffmann}, O.~B. and {Ilbert}, O. and {McCracken}, H.~J. and {Moneti}, A. and {Toft}, S. and {Brammer}, G. and {Shuntov}, M. and {Davidzon}, I. and {Hsieh}, B.~C. and {Laigle}, C. and {Anastasiou}, A. and {Jespersen}, C.~K. and {Vinther}, J. and {Capak}, P. and {Casey}, C.~M. and {McPartland}, C.~J.~R. and {Milvang-Jensen}, B. and {Mobasher}, B. and {Sanders}, D.~B. and {Zalesky}, L. and {Arnouts}, S. and {Aussel}, H. and {Dunlop}, J.~S. and {Faisst}, A. and {Franx}, M. and {Furtak}, L.~J. and {Fynbo}, J.~P.~U. and {Gould}, K.~M.~L. and {Greve}, T.~R. and {Gwyn}, S. and {Kartaltepe}, J.~S. and {Kashino}, D. and {Koekemoer}, A.~M. and {Kokorev}, V. and {Le F{\`e}vre}, O. and {Lilly}, S. and {Masters}, D. and {Magdis}, G. and {Mehta}, V. and {Peng}, Y. and {Riechers}, D.~A. and {Salvato}, M. and {Sawicki}, M. and {Scarlata}, C. and {Scoville}, N. and {Shirley}, R. and {Silverman}, J.~D. and {Sneppen}, A. and {Smolc̆i{\'c}}, V. and {Steinhardt}, C. and {Stern}, D. and {Tanaka}, M. and {Taniguchi}, Y. and {Teplitz}, H.~I. and {Vaccari}, M. and {Wang}, W. -H. and {Zamorani}, G.},
        title = "{COSMOS2020: A Panchromatic View of the Universe to z{\ensuremath{\sim}}10 from Two Complementary Catalogs}",
      journal = {\apjs},
         year = 2022,
        month = jan,
       volume = {258},
       number = {1},
          eid = {11},
        pages = {11},
          doi = {10.3847/1538-4365/ac3078},
archivePrefix = {arXiv},
       eprint = {2110.13923},
 primaryClass = {astro-ph.GA},
       adsurl = {https://ui.adsabs.harvard.edu/abs/2022ApJS..258...11W}
}

@ARTICLE{Kelson03,
       author = {{Kelson}, Daniel D.},
        title = "{Optimal Techniques in Two-dimensional Spectroscopy: Background Subtraction for the 21st Century}",
      journal = {\pasp},
         year = 2003,
        month = jun,
       volume = {115},
       number = {808},
        pages = {688-699},
          doi = {10.1086/375502},
archivePrefix = {arXiv},
       eprint = {astro-ph/0303507},
 primaryClass = {astro-ph},
       adsurl = {https://ui.adsabs.harvard.edu/abs/2003PASP..115..688K}
}

@ARTICLE{Kelson00,
       author = {{Kelson}, Daniel D. and {Illingworth}, Garth D. and {van Dokkum}, Pieter G. and {Franx}, Marijn},
        title = "{The Evolution of Early-Type Galaxies in Distant Clusters. II. Internal Kinematics of 55 Galaxies in the z=0.33 Cluster CL 1358+62}",
      journal = {\apj},
         year = 2000,
        month = mar,
       volume = {531},
       number = {1},
        pages = {159-183},
          doi = {10.1086/308445},
archivePrefix = {arXiv},
       eprint = {astro-ph/9908257},
 primaryClass = {astro-ph},
       adsurl = {https://ui.adsabs.harvard.edu/abs/2000ApJ...531..159K}
}

@ARTICLE{Scoville07,
       author = {{Scoville}, N. and {Aussel}, H. and {Brusa}, M. and {Capak}, P. and
         {Carollo}, C.~M. and {Elvis}, M. and {Giavalisco}, M. and {Guzzo}, L. and
         {Hasinger}, G. and {Impey}, C. and {Kneib}, J. -P. and {LeFevre}, O. and
         {Lilly}, S.~J. and {Mobasher}, B. and {Renzini}, A. and {Rich}, R.~M. and
         {Sanders}, D.~B. and {Schinnerer}, E. and {Schminovich}, D. and
         {Shopbell}, P. and {Taniguchi}, Y. and {Tyson}, N.~D.},
        title = "{The Cosmic Evolution Survey (COSMOS): Overview}",
      journal = {\apjs},
         year = 2007,
        month = sep,
       volume = {172},
       number = {1},
        pages = {1-8},
          doi = {10.1086/516585},
archivePrefix = {arXiv},
       eprint = {astro-ph/0612305},
 primaryClass = {astro-ph},
       adsurl = {https://ui.adsabs.harvard.edu/abs/2007ApJS..172....1S}
}

@ARTICLE{Muzzin13a,
       author = {{Muzzin}, Adam and {Marchesini}, Danilo and {Stefanon}, Mauro and {Franx}, Marijn and {Milvang-Jensen}, Bo and {Dunlop}, James S. and {Fynbo}, J.~P.~U. and {Brammer}, Gabriel and {Labb{\'e}}, Ivo and {van Dokkum}, Pieter},
        title = "{A Public K$_{s}$ -selected Catalog in the COSMOS/ULTRAVISTA Field: Photometry, Photometric Redshifts, and Stellar Population Parameters}",
      journal = {\apjs},
         year = 2013,
        month = may,
       volume = {206},
       number = {1},
          eid = {8},
        pages = {8},
          doi = {10.1088/0067-0049/206/1/8},
archivePrefix = {arXiv},
       eprint = {1303.4410},
 primaryClass = {astro-ph.CO},
       adsurl = {https://ui.adsabs.harvard.edu/abs/2013ApJS..206....8M}
}

@ARTICLE{Brammer08,
       author = {{Brammer}, Gabriel B. and {van Dokkum}, Pieter G. and {Coppi}, Paolo},
        title = "{EAZY: A Fast, Public Photometric Redshift Code}",
      journal = {\apj},
         year = 2008,
        month = oct,
       volume = {686},
       number = {2},
        pages = {1503-1513},
          doi = {10.1086/591786},
archivePrefix = {arXiv},
       eprint = {0807.1533},
 primaryClass = {astro-ph},
       adsurl = {https://ui.adsabs.harvard.edu/abs/2008ApJ...686.1503B}
}

@ARTICLE{Kriek09,
       author = {{Kriek}, Mariska and {van Dokkum}, Pieter G. and {Labb{\'e}}, Ivo and {Franx}, Marijn and {Illingworth}, Garth D. and {Marchesini}, Danilo and {Quadri}, Ryan F.},
        title = "{An Ultra-Deep Near-Infrared Spectrum of a Compact Quiescent Galaxy at z = 2.2}",
      journal = {\apj},
         year = 2009,
        month = jul,
       volume = {700},
       number = {1},
        pages = {221-231},
          doi = {10.1088/0004-637X/700/1/221},
archivePrefix = {arXiv},
       eprint = {0905.1692},
 primaryClass = {astro-ph.CO},
       adsurl = {https://ui.adsabs.harvard.edu/abs/2009ApJ...700..221K}
}

@ARTICLE{Arnouts02,
       author = {{Arnouts}, S. and {Moscardini}, L. and {Vanzella}, E. and {Colombi}, S. and {Cristiani}, S. and {Fontana}, A. and {Giallongo}, E. and {Matarrese}, S. and {Saracco}, P.},
        title = "{Measuring the redshift evolution of clustering: the Hubble Deep Field South}",
      journal = {\mnras},
         year = 2002,
        month = jan,
       volume = {329},
       number = {2},
        pages = {355-366},
          doi = {10.1046/j.1365-8711.2002.04988.x},
archivePrefix = {arXiv},
       eprint = {astro-ph/0109453},
 primaryClass = {astro-ph},
       adsurl = {https://ui.adsabs.harvard.edu/abs/2002MNRAS.329..355A}
}

@ARTICLE{Ilbert06,
       author = {{Ilbert}, O. and {Arnouts}, S. and {McCracken}, H.~J. and {Bolzonella}, M. and {Bertin}, E. and {Le F{\`e}vre}, O. and {Mellier}, Y. and {Zamorani}, G. and {Pell{\`o}}, R. and {Iovino}, A. and {Tresse}, L. and {Le Brun}, V. and {Bottini}, D. and {Garilli}, B. and {Maccagni}, D. and {Picat}, J.~P. and {Scaramella}, R. and {Scodeggio}, M. and {Vettolani}, G. and {Zanichelli}, A. and {Adami}, C. and {Bardelli}, S. and {Cappi}, A. and {Charlot}, S. and {Ciliegi}, P. and {Contini}, T. and {Cucciati}, O. and {Foucaud}, S. and {Franzetti}, P. and {Gavignaud}, I. and {Guzzo}, L. and {Marano}, B. and {Marinoni}, C. and {Mazure}, A. and {Meneux}, B. and {Merighi}, R. and {Paltani}, S. and {Pollo}, A. and {Pozzetti}, L. and {Radovich}, M. and {Zucca}, E. and {Bondi}, M. and {Bongiorno}, A. and {Busarello}, G. and {de La Torre}, S. and {Gregorini}, L. and {Lamareille}, F. and {Mathez}, G. and {Merluzzi}, P. and {Ripepi}, V. and {Rizzo}, D. and {Vergani}, D.},
        title = "{Accurate photometric redshifts for the CFHT legacy survey calibrated using the VIMOS VLT deep survey}",
      journal = {\aap},
         year = 2006,
        month = oct,
       volume = {457},
       number = {3},
        pages = {841-856},
          doi = {10.1051/0004-6361:20065138},
archivePrefix = {arXiv},
       eprint = {astro-ph/0603217},
 primaryClass = {astro-ph},
       adsurl = {https://ui.adsabs.harvard.edu/abs/2006A&A...457..841I}
}

@ARTICLE{Horne86,
       author = {{Horne}, K.},
        title = "{An optimal extraction algorithm for CCD spectroscopy.}",
      journal = {\pasp},
         year = 1986,
        month = jun,
       volume = {98},
        pages = {609-617},
          doi = {10.1086/131801},
       adsurl = {https://ui.adsabs.harvard.edu/abs/1986PASP...98..609H}
}

@ARTICLE{Carnall18,
       author = {{Carnall}, A.~C. and {McLure}, R.~J. and {Dunlop}, J.~S. and
         {Dav{\'e}}, R.},
        title = "{Inferring the star formation histories of massive quiescent galaxies with BAGPIPES: evidence for multiple quenching mechanisms}",
      journal = {\mnras},
         year = 2018,
        month = nov,
       volume = {480},
       number = {4},
        pages = {4379-4401},
          doi = {10.1093/mnras/sty2169},
archivePrefix = {arXiv},
       eprint = {1712.04452},
 primaryClass = {astro-ph.GA},
       adsurl = {https://ui.adsabs.harvard.edu/abs/2018MNRAS.480.4379C}
}

@ARTICLE{Byler17,
       author = {{Byler}, Nell and {Dalcanton}, Julianne J. and {Conroy}, Charlie and
         {Johnson}, Benjamin D.},
        title = "{Nebular Continuum and Line Emission in Stellar Population Synthesis Models}",
      journal = {\apj},
         year = 2017,
        month = may,
       volume = {840},
       number = {1},
          eid = {44},
        pages = {44},
          doi = {10.3847/1538-4357/aa6c66},
archivePrefix = {arXiv},
       eprint = {1611.08305},
 primaryClass = {astro-ph.GA},
       adsurl = {https://ui.adsabs.harvard.edu/abs/2017ApJ...840...44B}
}

@ARTICLE{Bruzual03,
       author = {{Bruzual}, G. and {Charlot}, S.},
        title = "{Stellar population synthesis at the resolution of 2003}",
      journal = {\mnras},
         year = 2003,
        month = oct,
       volume = {344},
       number = {4},
        pages = {1000-1028},
          doi = {10.1046/j.1365-8711.2003.06897.x},
archivePrefix = {arXiv},
       eprint = {astro-ph/0309134},
 primaryClass = {astro-ph},
       adsurl = {https://ui.adsabs.harvard.edu/abs/2003MNRAS.344.1000B}
}

@ARTICLE{Calzetti00,
       author = {{Calzetti}, Daniela and {Armus}, Lee and {Bohlin}, Ralph C. and
         {Kinney}, Anne L. and {Koornneef}, Jan and {Storchi-Bergmann}, Thaisa},
        title = "{The Dust Content and Opacity of Actively Star-forming Galaxies}",
      journal = {\apj},
         year = 2000,
        month = apr,
       volume = {533},
       number = {2},
        pages = {682-695},
          doi = {10.1086/308692},
archivePrefix = {arXiv},
       eprint = {astro-ph/9911459},
 primaryClass = {astro-ph},
       adsurl = {https://ui.adsabs.harvard.edu/abs/2000ApJ...533..682C}
}

@ARTICLE{Ferland17,
       author = {{Ferland}, G.~J. and {Chatzikos}, M. and {Guzm{\'a}n}, F. and {Lykins}, M.~L. and {van Hoof}, P.~A.~M. and {Williams}, R.~J.~R. and {Abel}, N.~P. and {Badnell}, N.~R. and {Keenan}, F.~P. and {Porter}, R.~L. and {Stancil}, P.~C.},
        title = "{The 2017 Release Cloudy}",
      journal = {\rmxaa},
         year = 2017,
        month = oct,
       volume = {53},
        pages = {385-438},
          doi = {10.48550/arXiv.1705.10877},
archivePrefix = {arXiv},
       eprint = {1705.10877},
 primaryClass = {astro-ph.GA},
       adsurl = {https://ui.adsabs.harvard.edu/abs/2017RMxAA..53..385F}
}

@ARTICLE{Kroupa01,
       author = {{Kroupa}, Pavel},
        title = "{On the variation of the initial mass function}",
      journal = {\mnras},
         year = 2001,
        month = apr,
       volume = {322},
       number = {2},
        pages = {231-246},
          doi = {10.1046/j.1365-8711.2001.04022.x},
archivePrefix = {arXiv},
       eprint = {astro-ph/0009005},
 primaryClass = {astro-ph},
       adsurl = {https://ui.adsabs.harvard.edu/abs/2001MNRAS.322..231K}
}

@ARTICLE{Muzzin13b,
       author = {{Muzzin}, Adam and {Marchesini}, Danilo and {Stefanon}, Mauro and {Franx}, Marijn and {McCracken}, Henry J. and {Milvang-Jensen}, Bo and {Dunlop}, James S. and {Fynbo}, J.~P.~U. and {Brammer}, Gabriel and {Labb{\'e}}, Ivo and {van Dokkum}, Pieter G.},
        title = "{The Evolution of the Stellar Mass Functions of Star-forming and Quiescent Galaxies to z = 4 from the COSMOS/UltraVISTA Survey}",
      journal = {\apj},
         year = 2013,
        month = nov,
       volume = {777},
       number = {1},
          eid = {18},
        pages = {18},
          doi = {10.1088/0004-637X/777/1/18},
archivePrefix = {arXiv},
       eprint = {1303.4409},
 primaryClass = {astro-ph.CO},
       adsurl = {https://ui.adsabs.harvard.edu/abs/2013ApJ...777...18M}
}

@BOOK{Osterbrock06,
       author = {{Osterbrock}, Donald E. and {Ferland}, Gary J.},
        title = "{Astrophysics of gaseous nebulae and active galactic nuclei}",
         year = 2006,
       adsurl = {https://ui.adsabs.harvard.edu/abs/2006agna.book.....O}
}

@ARTICLE{Garn10,
       author = {{Garn}, Timothy and {Best}, Philip N.},
        title = "{Predicting dust extinction from the stellar mass of a galaxy}",
      journal = {\mnras},
         year = 2010,
        month = nov,
       volume = {409},
       number = {1},
        pages = {421-432},
          doi = {10.1111/j.1365-2966.2010.17321.x},
archivePrefix = {arXiv},
       eprint = {1007.1145},
 primaryClass = {astro-ph.GA},
       adsurl = {https://ui.adsabs.harvard.edu/abs/2010MNRAS.409..421G}
}

@ARTICLE{Charlot00,
       author = {{Charlot}, St{\'e}phane and {Fall}, S. Michael},
        title = "{A Simple Model for the Absorption of Starlight by Dust in Galaxies}",
      journal = {\apj},
         year = 2000,
        month = aug,
       volume = {539},
       number = {2},
        pages = {718-731},
          doi = {10.1086/309250},
archivePrefix = {arXiv},
       eprint = {astro-ph/0003128},
 primaryClass = {astro-ph},
       adsurl = {https://ui.adsabs.harvard.edu/abs/2000ApJ...539..718C}
}

@ARTICLE{Cullen18,
       author = {{Cullen}, F. and {McLure}, R.~J. and {Khochfar}, S. and {Dunlop}, J.~S. and {Dalla Vecchia}, C. and {Carnall}, A.~C. and {Bourne}, N. and {Castellano}, M. and {Cimatti}, A. and {Cirasuolo}, M. and {Elbaz}, D. and {Fynbo}, J.~P.~U. and {Garilli}, B. and {Koekemoer}, A. and {Marchi}, F. and {Pentericci}, L. and {Talia}, M. and {Zamorani}, G.},
        title = "{The VANDELS survey: dust attenuation in star-forming galaxies at z = 3-4}",
      journal = {\mnras},
         year = 2018,
        month = may,
       volume = {476},
       number = {3},
        pages = {3218-3232},
          doi = {10.1093/mnras/sty469},
archivePrefix = {arXiv},
       eprint = {1712.01292},
 primaryClass = {astro-ph.GA},
       adsurl = {https://ui.adsabs.harvard.edu/abs/2018MNRAS.476.3218C}
}

@ARTICLE{Shivaei20,
       author = {{Shivaei}, Irene and {Reddy}, Naveen and {Rieke}, George and {Shapley}, Alice and {Kriek}, Mariska and {Battisti}, Andrew and {Mobasher}, Bahram and {Sanders}, Ryan and {Fetherolf}, Tara and {Azadi}, Mojegan and {Coil}, Alison L. and {Freeman}, William R. and {de Groot}, Laura and {Leung}, Gene and {Price}, Sedona H. and {Siana}, Brian and {Zick}, Tom},
        title = "{The MOSDEF Survey: The Variation of the Dust Attenuation Curve with Metallicity}",
      journal = {\apj},
         year = 2020,
        month = aug,
       volume = {899},
       number = {2},
          eid = {117},
        pages = {117},
          doi = {10.3847/1538-4357/aba35e},
archivePrefix = {arXiv},
       eprint = {2005.01742},
 primaryClass = {astro-ph.GA},
       adsurl = {https://ui.adsabs.harvard.edu/abs/2020ApJ...899..117S}
}

@ARTICLE{Hemmati15,
       author = {{Hemmati}, Shoubaneh and {Mobasher}, Bahram and {Darvish}, Behnam and {Nayyeri}, Hooshang and {Sobral}, David and {Miller}, Sarah},
        title = "{Nebular and Stellar Dust Extinction Across the Disk of Emission-line Galaxies on Kiloparsec Scales}",
      journal = {\apj},
         year = 2015,
        month = nov,
       volume = {814},
       number = {1},
          eid = {46},
        pages = {46},
          doi = {10.1088/0004-637X/814/1/46},
archivePrefix = {arXiv},
       eprint = {1510.02506},
 primaryClass = {astro-ph.GA},
       adsurl = {https://ui.adsabs.harvard.edu/abs/2015ApJ...814...46H}
}

@ARTICLE{Chartab24,
       author = {{Chartab}, Nima and {Newman}, Andrew B. and {Rudie}, Gwen C. and {Blanc}, Guillermo A. and {Kelson}, Daniel D.},
        title = "{LATIS: The Stellar Mass-Metallicity Relation of Star-forming Galaxies at z   2.5}",
      journal = {\apj},
         year = 2024,
        month = jan,
       volume = {960},
       number = {1},
          eid = {73},
        pages = {73},
          doi = {10.3847/1538-4357/ad0554},
archivePrefix = {arXiv},
       eprint = {2310.12200},
 primaryClass = {astro-ph.GA},
       adsurl = {https://ui.adsabs.harvard.edu/abs/2024ApJ...960...73C}
}

@ARTICLE{Qin19,
       author = {{Qin}, Jianbo and {Zheng}, Xian Zhong and {Wuyts}, Stijn and {Pan}, Zhizheng and {Ren}, Jian},
        title = "{Understanding the Discrepancy between IRX and Balmer Decrement in Tracing Galaxy Dust Attenuation}",
      journal = {\apj},
         year = 2019,
        month = nov,
       volume = {886},
       number = {1},
          eid = {28},
        pages = {28},
          doi = {10.3847/1538-4357/ab4a04},
archivePrefix = {arXiv},
       eprint = {1909.13505},
 primaryClass = {astro-ph.GA},
       adsurl = {https://ui.adsabs.harvard.edu/abs/2019ApJ...886...28Q}
}

@ARTICLE{Reddy15,
       author = {{Reddy}, Naveen A. and {Kriek}, Mariska and {Shapley}, Alice E. and {Freeman}, William R. and {Siana}, Brian and {Coil}, Alison L. and {Mobasher}, Bahram and {Price}, Sedona H. and {Sanders}, Ryan L. and {Shivaei}, Irene},
        title = "{The MOSDEF Survey: Measurements of Balmer Decrements and the Dust Attenuation Curve at Redshifts z \raisebox{-0.5ex}\textasciitilde 1.4-2.6}",
      journal = {\apj},
         year = 2015,
        month = jun,
       volume = {806},
       number = {2},
          eid = {259},
        pages = {259},
          doi = {10.1088/0004-637X/806/2/259},
archivePrefix = {arXiv},
       eprint = {1504.02782},
 primaryClass = {astro-ph.GA},
       adsurl = {https://ui.adsabs.harvard.edu/abs/2015ApJ...806..259R}
}

@ARTICLE{Kennicutt98,
       author = {{Kennicutt}, Robert C., Jr.},
        title = "{Star Formation in Galaxies Along the Hubble Sequence}",
      journal = {\araa},
         year = 1998,
        month = jan,
       volume = {36},
        pages = {189-232},
          doi = {10.1146/annurev.astro.36.1.189},
archivePrefix = {arXiv},
       eprint = {astro-ph/9807187},
 primaryClass = {astro-ph},
       adsurl = {https://ui.adsabs.harvard.edu/abs/1998ARA&A..36..189K}
}

@ARTICLE{White78,
       author = {{White}, S.~D.~M. and {Rees}, M.~J.},
        title = "{Core condensation in heavy halos: a two-stage theory for galaxy formation and clustering.}",
      journal = {\mnras},
         year = 1978,
        month = may,
       volume = {183},
        pages = {341-358},
          doi = {10.1093/mnras/183.3.341},
       adsurl = {https://ui.adsabs.harvard.edu/abs/1978MNRAS.183..341W}
}

@ARTICLE{Dekel86,
       author = {{Dekel}, A. and {Silk}, J.},
        title = "{The Origin of Dwarf Galaxies, Cold Dark Matter, and Biased Galaxy Formation}",
      journal = {\apj},
         year = 1986,
        month = apr,
       volume = {303},
        pages = {39},
          doi = {10.1086/164050},
       adsurl = {https://ui.adsabs.harvard.edu/abs/1986ApJ...303...39D}
}

@ARTICLE{Tremonti04,
       author = {{Tremonti}, Christy A. and {Heckman}, Timothy M. and {Kauffmann}, Guinevere and {Brinchmann}, Jarle and {Charlot}, St{\'e}phane and {White}, Simon D.~M. and {Seibert}, Mark and {Peng}, Eric W. and {Schlegel}, David J. and {Uomoto}, Alan and {Fukugita}, Masataka and {Brinkmann}, Jon},
        title = "{The Origin of the Mass-Metallicity Relation: Insights from 53,000 Star-forming Galaxies in the Sloan Digital Sky Survey}",
      journal = {\apj},
         year = 2004,
        month = oct,
       volume = {613},
       number = {2},
        pages = {898-913},
          doi = {10.1086/423264},
archivePrefix = {arXiv},
       eprint = {astro-ph/0405537},
 primaryClass = {astro-ph},
       adsurl = {https://ui.adsabs.harvard.edu/abs/2004ApJ...613..898T}
}

@ARTICLE{Peeples11,
       author = {{Peeples}, Molly S. and {Shankar}, Francesco},
        title = "{Constraints on star formation driven galaxy winds from the mass-metallicity relation at z= 0}",
      journal = {\mnras},
         year = 2011,
        month = nov,
       volume = {417},
       number = {4},
        pages = {2962-2981},
          doi = {10.1111/j.1365-2966.2011.19456.x},
archivePrefix = {arXiv},
       eprint = {1007.3743},
 primaryClass = {astro-ph.CO},
       adsurl = {https://ui.adsabs.harvard.edu/abs/2011MNRAS.417.2962P}
}

@ARTICLE{Baldry12,
       author = {{Baldry}, I.~K. and {Driver}, S.~P. and {Loveday}, J. and {Taylor}, E.~N. and {Kelvin}, L.~S. and {Liske}, J. and {Norberg}, P. and {Robotham}, A.~S.~G. and {Brough}, S. and {Hopkins}, A.~M. and {Bamford}, S.~P. and {Peacock}, J.~A. and {Bland-Hawthorn}, J. and {Conselice}, C.~J. and {Croom}, S.~M. and {Jones}, D.~H. and {Parkinson}, H.~R. and {Popescu}, C.~C. and {Prescott}, M. and {Sharp}, R.~G. and {Tuffs}, R.~J.},
        title = "{Galaxy And Mass Assembly (GAMA): the galaxy stellar mass function at z < 0.06}",
      journal = {\mnras},
         year = 2012,
        month = mar,
       volume = {421},
       number = {1},
        pages = {621-634},
          doi = {10.1111/j.1365-2966.2012.20340.x},
archivePrefix = {arXiv},
       eprint = {1111.5707},
 primaryClass = {astro-ph.CO},
       adsurl = {https://ui.adsabs.harvard.edu/abs/2012MNRAS.421..621B}
}

@ARTICLE{Lin23,
       author = {{Lin}, Xiaojing and {Cai}, Zheng and {Zou}, Siwei and {Li}, Zihao and {Chen}, Zuyi and {Bian}, Fuyan and {Sun}, Fengwu and {Shu}, Yiping and {Wu}, Yunjing and {Li}, Mingyu and {Li}, Jianan and {Fan}, Xiaohui and {Prochaska}, J. Xavier and {Schaerer}, Daniel and {Charlot}, Stephane and {Espada}, Daniel and {Dessauges-Zavadsky}, Miroslava and {Egami}, Eiichi and {Stark}, Daniel and {Knudsen}, Kirsten K. and {Bruzual}, Gustavo and {Chevallard}, Jacopo},
        title = "{Metal-enriched Neutral Gas Reservoir around a Strongly Lensed Low-mass Galaxy at z = 4 Identified by JWST/NIRISS and VLT/MUSE}",
      journal = {\apjl},
         year = 2023,
        month = feb,
       volume = {944},
       number = {2},
          eid = {L59},
        pages = {L59},
          doi = {10.3847/2041-8213/aca1c4},
archivePrefix = {arXiv},
       eprint = {2209.03376},
 primaryClass = {astro-ph.GA},
       adsurl = {https://ui.adsabs.harvard.edu/abs/2023ApJ...944L..59L}
}

@ARTICLE{Robertson10,
       author = {{Robertson}, Brant E. and {Ellis}, Richard S. and {Dunlop}, James S. and {McLure}, Ross J. and {Stark}, Daniel P.},
        title = "{Early star-forming galaxies and the reionization of the Universe}",
      journal = {\nat},
         year = 2010,
        month = nov,
       volume = {468},
       number = {7320},
        pages = {49-55},
          doi = {10.1038/nature09527},
archivePrefix = {arXiv},
       eprint = {1011.0727},
 primaryClass = {astro-ph.CO},
       adsurl = {https://ui.adsabs.harvard.edu/abs/2010Natur.468...49R}
}

@ARTICLE{Kashino13,
       author = {{Kashino}, D. and {Silverman}, J.~D. and {Rodighiero}, G. and {Renzini}, A. and {Arimoto}, N. and {Daddi}, E. and {Lilly}, S.~J. and {Sanders}, D.~B. and {Kartaltepe}, J. and {Zahid}, H.~J. and {Nagao}, T. and {Sugiyama}, N. and {Capak}, P. and {Carollo}, C.~M. and {Chu}, J. and {Hasinger}, G. and {Ilbert}, O. and {Kajisawa}, M. and {Kewley}, L.~J. and {Koekemoer}, A.~M. and {Kova{\v{c}}}, K. and {Le F{\`e}vre}, O. and {Masters}, D. and {McCracken}, H.~J. and {Onodera}, M. and {Scoville}, N. and {Strazzullo}, V. and {Symeonidis}, M. and {Taniguchi}, Y.},
        title = "{The FMOS-COSMOS Survey of Star-forming Galaxies at z \raisebox{-0.5ex}\textasciitilde 1.6. I. H{\ensuremath{\alpha}}-based Star Formation Rates and Dust Extinction}",
      journal = {\apjl},
         year = 2013,
        month = nov,
       volume = {777},
       number = {1},
          eid = {L8},
        pages = {L8},
          doi = {10.1088/2041-8205/777/1/L8},
archivePrefix = {arXiv},
       eprint = {1309.4774},
 primaryClass = {astro-ph.CO},
       adsurl = {https://ui.adsabs.harvard.edu/abs/2013ApJ...777L...8K}
}

@ARTICLE{Speagle14,
       author = {{Speagle}, J.~S. and {Steinhardt}, C.~L. and {Capak}, P.~L. and {Silverman}, J.~D.},
        title = "{A Highly Consistent Framework for the Evolution of the Star-Forming ``Main Sequence'' from z \raisebox{-0.5ex}\textasciitilde 0-6}",
      journal = {\apjs},
         year = 2014,
        month = oct,
       volume = {214},
       number = {2},
          eid = {15},
        pages = {15},
          doi = {10.1088/0067-0049/214/2/15},
archivePrefix = {arXiv},
       eprint = {1405.2041},
 primaryClass = {astro-ph.GA},
       adsurl = {https://ui.adsabs.harvard.edu/abs/2014ApJS..214...15S}
}

@ARTICLE{Kelson14,
       author = {{Kelson}, Daniel D.},
        title = "{Decoding the Star-Forming Main Sequence or: How I Learned to Stop Worrying and Love the Central Limit Theorem}",
      journal = {arXiv e-prints},
         year = 2014,
        month = jun,
          eid = {arXiv:1406.5191},
        pages = {arXiv:1406.5191},
          doi = {10.48550/arXiv.1406.5191},
archivePrefix = {arXiv},
       eprint = {1406.5191},
 primaryClass = {astro-ph.GA},
       adsurl = {https://ui.adsabs.harvard.edu/abs/2014arXiv1406.5191K}
}

@ARTICLE{Whitaker14,
       author = {{Whitaker}, Katherine E. and {Franx}, Marijn and {Leja}, Joel and {van Dokkum}, Pieter G. and {Henry}, Alaina and {Skelton}, Rosalind E. and {Fumagalli}, Mattia and {Momcheva}, Ivelina G. and {Brammer}, Gabriel B. and {Labb{\'e}}, Ivo and {Nelson}, Erica J. and {Rigby}, Jane R.},
        title = "{Constraining the Low-mass Slope of the Star Formation Sequence at 0.5 < z < 2.5}",
      journal = {\apj},
         year = 2014,
        month = nov,
       volume = {795},
       number = {2},
          eid = {104},
        pages = {104},
          doi = {10.1088/0004-637X/795/2/104},
archivePrefix = {arXiv},
       eprint = {1407.1843},
 primaryClass = {astro-ph.GA},
       adsurl = {https://ui.adsabs.harvard.edu/abs/2014ApJ...795..104W}
}

@ARTICLE{Salmon16,
       author = {{Salmon}, Brett and {Papovich}, Casey and {Long}, James and {Willner}, S.~P. and {Finkelstein}, Steven L. and {Ferguson}, Henry C. and {Dickinson}, Mark and {Duncan}, Kenneth and {Faber}, S.~M. and {Hathi}, Nimish and {Koekemoer}, Anton and {Kurczynski}, Peter and {Newman}, Jeffery and {Pacifici}, Camilla and {P{\'e}rez-Gonz{\'a}lez}, Pablo G. and {Pforr}, Janine},
        title = "{Breaking the Curve with CANDELS: A Bayesian Approach to Reveal the Non-Universality of the Dust-Attenuation Law at High Redshift}",
      journal = {\apj},
         year = 2016,
        month = aug,
       volume = {827},
       number = {1},
          eid = {20},
        pages = {20},
          doi = {10.3847/0004-637X/827/1/20},
archivePrefix = {arXiv},
       eprint = {1512.05396},
 primaryClass = {astro-ph.GA},
       adsurl = {https://ui.adsabs.harvard.edu/abs/2016ApJ...827...20S}
}

@ARTICLE{Reddy18,
       author = {{Reddy}, Naveen A. and {Oesch}, Pascal A. and {Bouwens}, Rychard J. and {Montes}, Mireia and {Illingworth}, Garth D. and {Steidel}, Charles C. and {van Dokkum}, Pieter G. and {Atek}, Hakim and {Carollo}, Marcella C. and {Cibinel}, Anna and {Holden}, Brad and {Labb{\'e}}, Ivo and {Magee}, Dan and {Morselli}, Laura and {Nelson}, Erica J. and {Wilkins}, Steve},
        title = "{The HDUV Survey: A Revised Assessment of the Relationship between UV Slope and Dust Attenuation for High-redshift Galaxies}",
      journal = {\apj},
         year = 2018,
        month = jan,
       volume = {853},
       number = {1},
          eid = {56},
        pages = {56},
          doi = {10.3847/1538-4357/aaa3e7},
archivePrefix = {arXiv},
       eprint = {1705.09302},
 primaryClass = {astro-ph.GA},
       adsurl = {https://ui.adsabs.harvard.edu/abs/2018ApJ...853...56R}
}

@ARTICLE{Gordon03,
       author = {{Gordon}, Karl D. and {Clayton}, Geoffrey C. and {Misselt}, K.~A. and {Landolt}, Arlo U. and {Wolff}, Michael J.},
        title = "{A Quantitative Comparison of the Small Magellanic Cloud, Large Magellanic Cloud, and Milky Way Ultraviolet to Near-Infrared Extinction Curves}",
      journal = {\apj},
         year = 2003,
        month = sep,
       volume = {594},
       number = {1},
        pages = {279-293},
          doi = {10.1086/376774},
archivePrefix = {arXiv},
       eprint = {astro-ph/0305257},
 primaryClass = {astro-ph},
       adsurl = {https://ui.adsabs.harvard.edu/abs/2003ApJ...594..279G}
}

@ARTICLE{Salim18,
       author = {{Salim}, Samir and {Boquien}, M{\'e}d{\'e}ric and {Lee}, Janice C.},
        title = "{Dust Attenuation Curves in the Local Universe: Demographics and New Laws for Star-forming Galaxies and High-redshift Analogs}",
      journal = {\apj},
         year = 2018,
        month = may,
       volume = {859},
       number = {1},
          eid = {11},
        pages = {11},
          doi = {10.3847/1538-4357/aabf3c},
archivePrefix = {arXiv},
       eprint = {1804.05850},
 primaryClass = {astro-ph.GA},
       adsurl = {https://ui.adsabs.harvard.edu/abs/2018ApJ...859...11S}
}

@ARTICLE{Weisz12,
       author = {{Weisz}, Daniel R. and {Johnson}, Benjamin D. and {Johnson}, L. Clifton and {Skillman}, Evan D. and {Lee}, Janice C. and {Kennicutt}, Robert C. and {Calzetti}, Daniela and {van Zee}, Liese and {Bothwell}, Matthew S. and {Dalcanton}, Julianne J. and {Dale}, Daniel A. and {Williams}, Benjamin F.},
        title = "{Modeling the Effects of Star Formation Histories on H{\ensuremath{\alpha}} and Ultraviolet Fluxes in nearby Dwarf Galaxies}",
      journal = {\apj},
         year = 2012,
        month = jan,
       volume = {744},
       number = {1},
          eid = {44},
        pages = {44},
          doi = {10.1088/0004-637X/744/1/44},
archivePrefix = {arXiv},
       eprint = {1109.2905},
 primaryClass = {astro-ph.CO},
       adsurl = {https://ui.adsabs.harvard.edu/abs/2012ApJ...744...44W}
}

@ARTICLE{Boogaard18,
       author = {{Boogaard}, Leindert A. and {Brinchmann}, Jarle and {Bouch{\'e}}, Nicolas and {Paalvast}, Mieke and {Bacon}, Roland and {Bouwens}, Rychard J. and {Contini}, Thierry and {Gunawardhana}, Madusha L.~P. and {Inami}, Hanae and {Marino}, Raffaella A. and {Maseda}, Michael V. and {Mitchell}, Peter and {Nanayakkara}, Themiya and {Richard}, Johan and {Schaye}, Joop and {Schreiber}, Corentin and {Tacchella}, Sandro and {Wisotzki}, Lutz and {Zabl}, Johannes},
        title = "{The MUSE Hubble Ultra Deep Field Survey. XI. Constraining the low-mass end of the stellar mass - star formation rate relation at z < 1}",
      journal = {\aap},
         year = 2018,
        month = nov,
       volume = {619},
          eid = {A27},
        pages = {A27},
          doi = {10.1051/0004-6361/201833136},
archivePrefix = {arXiv},
       eprint = {1808.04900},
 primaryClass = {astro-ph.GA},
       adsurl = {https://ui.adsabs.harvard.edu/abs/2018A&A...619A..27B}
}

@ARTICLE{Emami19,
       author = {{Emami}, Najmeh and {Siana}, Brian and {Weisz}, Daniel R. and {Johnson}, Benjamin D. and {Ma}, Xiangcheng and {El-Badry}, Kareem},
        title = "{A Closer Look at Bursty Star Formation with L $_{H{\ensuremath{\alpha}} }$ and L $_{UV}$ Distributions}",
      journal = {\apj},
         year = 2019,
        month = aug,
       volume = {881},
       number = {1},
          eid = {71},
        pages = {71},
          doi = {10.3847/1538-4357/ab211a},
archivePrefix = {arXiv},
       eprint = {1809.06380},
 primaryClass = {astro-ph.GA},
       adsurl = {https://ui.adsabs.harvard.edu/abs/2019ApJ...881...71E}
}

@ARTICLE{Noeske07,
       author = {{Noeske}, K.~G. and {Weiner}, B.~J. and {Faber}, S.~M. and {Papovich}, C. and {Koo}, D.~C. and {Somerville}, R.~S. and {Bundy}, K. and {Conselice}, C.~J. and {Newman}, J.~A. and {Schiminovich}, D. and {Le Floc'h}, E. and {Coil}, A.~L. and {Rieke}, G.~H. and {Lotz}, J.~M. and {Primack}, J.~R. and {Barmby}, P. and {Cooper}, M.~C. and {Davis}, M. and {Ellis}, R.~S. and {Fazio}, G.~G. and {Guhathakurta}, P. and {Huang}, J. and {Kassin}, S.~A. and {Martin}, D.~C. and {Phillips}, A.~C. and {Rich}, R.~M. and {Small}, T.~A. and {Willmer}, C.~N.~A. and {Wilson}, G.},
        title = "{Star Formation in AEGIS Field Galaxies since z=1.1: The Dominance of Gradually Declining Star Formation, and the Main Sequence of Star-forming Galaxies}",
      journal = {\apjl},
         year = 2007,
        month = may,
       volume = {660},
       number = {1},
        pages = {L43-L46},
          doi = {10.1086/517926},
archivePrefix = {arXiv},
       eprint = {astro-ph/0701924},
 primaryClass = {astro-ph},
       adsurl = {https://ui.adsabs.harvard.edu/abs/2007ApJ...660L..43N}
}

@ARTICLE{Salim20,
       author = {{Salim}, Samir and {Narayanan}, Desika},
        title = "{The Dust Attenuation Law in Galaxies}",
      journal = {\araa},
         year = 2020,
        month = aug,
       volume = {58},
        pages = {529-575},
          doi = {10.1146/annurev-astro-032620-021933},
archivePrefix = {arXiv},
       eprint = {2001.03181},
 primaryClass = {astro-ph.GA},
       adsurl = {https://ui.adsabs.harvard.edu/abs/2020ARA&A..58..529S}
}

@ARTICLE{RemyRuyer14,
       author = {{R{\'e}my-Ruyer}, A. and {Madden}, S.~C. and {Galliano}, F. and {Galametz}, M. and {Takeuchi}, T.~T. and {Asano}, R.~S. and {Zhukovska}, S. and {Lebouteiller}, V. and {Cormier}, D. and {Jones}, A. and {Bocchio}, M. and {Baes}, M. and {Bendo}, G.~J. and {Boquien}, M. and {Boselli}, A. and {DeLooze}, I. and {Doublier-Pritchard}, V. and {Hughes}, T. and {Karczewski}, O. {\L}. and {Spinoglio}, L.},
        title = "{Gas-to-dust mass ratios in local galaxies over a 2 dex metallicity range}",
      journal = {\aap},
         year = 2014,
        month = mar,
       volume = {563},
          eid = {A31},
        pages = {A31},
          doi = {10.1051/0004-6361/201322803},
archivePrefix = {arXiv},
       eprint = {1312.3442},
 primaryClass = {astro-ph.GA},
       adsurl = {https://ui.adsabs.harvard.edu/abs/2014A&A...563A..31R}
}

@ARTICLE{DeVis19,
       author = {{De Vis}, P. and {Jones}, A. and {Viaene}, S. and {Casasola}, V. and {Clark}, C.~J.~R. and {Baes}, M. and {Bianchi}, S. and {Cassara}, L.~P. and {Davies}, J.~I. and {De Looze}, I. and {Galametz}, M. and {Galliano}, F. and {Lianou}, S. and {Madden}, S. and {Manilla-Robles}, A. and {Mosenkov}, A.~V. and {Nersesian}, A. and {Roychowdhury}, S. and {Xilouris}, E.~M. and {Ysard}, N.},
        title = "{A systematic metallicity study of DustPedia galaxies reveals evolution in the dust-to-metal ratios}",
      journal = {\aap},
         year = 2019,
        month = mar,
       volume = {623},
          eid = {A5},
        pages = {A5},
          doi = {10.1051/0004-6361/201834444},
archivePrefix = {arXiv},
       eprint = {1901.09040},
 primaryClass = {astro-ph.GA},
       adsurl = {https://ui.adsabs.harvard.edu/abs/2019A&A...623A...5D}
}

@ARTICLE{Price14,
       author = {{Price}, Sedona H. and {Kriek}, Mariska and {Brammer}, Gabriel B. and {Conroy}, Charlie and {F{\"o}rster Schreiber}, Natascha M. and {Franx}, Marijn and {Fumagalli}, Mattia and {Lundgren}, Britt and {Momcheva}, Ivelina and {Nelson}, Erica J. and {Skelton}, Rosalind E. and {van Dokkum}, Pieter G. and {Whitaker}, Katherine E. and {Wuyts}, Stijn},
        title = "{Direct Measurements of Dust Attenuation in z \raisebox{-0.5ex}\textasciitilde 1.5 Star-forming Galaxies from 3D-HST: Implications for Dust Geometry and Star Formation Rates}",
      journal = {\apj},
         year = 2014,
        month = jun,
       volume = {788},
       number = {1},
          eid = {86},
        pages = {86},
          doi = {10.1088/0004-637X/788/1/86},
archivePrefix = {arXiv},
       eprint = {1310.4177},
 primaryClass = {astro-ph.CO},
       adsurl = {https://ui.adsabs.harvard.edu/abs/2014ApJ...788...86P}
}

@ARTICLE{Kelson20,
       author = {{Kelson}, Daniel D. and {Abramson}, Louis E. and {Benson}, Andrew J. and {Patel}, Shannon G. and {Shectman}, Stephen A. and {Dressler}, Alan and {McCarthy}, Patrick J. and {Mulchaey}, John S. and {Williams}, Rik J.},
        title = "{Gravity and the non-linear growth of structure in the Carnegie-Spitzer-IMACS Redshift Survey}",
      journal = {\mnras},
         year = 2020,
        month = may,
       volume = {494},
       number = {2},
        pages = {2628-2640},
          doi = {10.1093/mnras/staa100},
archivePrefix = {arXiv},
       eprint = {1908.08952},
 primaryClass = {astro-ph.CO},
       adsurl = {https://ui.adsabs.harvard.edu/abs/2020MNRAS.494.2628K}
}

@ARTICLE{Weisz14,
       author = {{Weisz}, Daniel R. and {Dolphin}, Andrew E. and {Skillman}, Evan D. and {Holtzman}, Jon and {Gilbert}, Karoline M. and {Dalcanton}, Julianne J. and {Williams}, Benjamin F.},
        title = "{The Star Formation Histories of Local Group Dwarf Galaxies. II. Searching For Signatures of Reionization}",
      journal = {\apj},
         year = 2014,
        month = jul,
       volume = {789},
       number = {2},
          eid = {148},
        pages = {148},
          doi = {10.1088/0004-637X/789/2/148},
archivePrefix = {arXiv},
       eprint = {1405.3281},
 primaryClass = {astro-ph.GA},
       adsurl = {https://ui.adsabs.harvard.edu/abs/2014ApJ...789..148W}
}

@ARTICLE{Lee09,
       author = {{Lee}, Janice C. and {Gil de Paz}, Armando and {Tremonti}, Christy and {Kennicutt}, Jr., Robert C. and {Salim}, Samir and {Bothwell}, Matthew and {Calzetti}, Daniela and {Dalcanton}, Julianne and {Dale}, Daniel and {Engelbracht}, Chad and {Funes}, S.~J. Jos{\'e} G. and {Johnson}, Benjamin and {Sakai}, Shoko and {Skillman}, Evan and {van Zee}, Liese and {Walter}, Fabian and {Weisz}, Daniel},
        title = "{Comparison of H{\ensuremath{\alpha}} and UV Star Formation Rates in the Local Volume: Systematic Discrepancies for Dwarf Galaxies}",
      journal = {\apj},
         year = 2009,
        month = nov,
       volume = {706},
       number = {1},
        pages = {599-613},
          doi = {10.1088/0004-637X/706/1/599},
archivePrefix = {arXiv},
       eprint = {0909.5205},
 primaryClass = {astro-ph.CO},
       adsurl = {https://ui.adsabs.harvard.edu/abs/2009ApJ...706..599L}
}

@ARTICLE{Gallagher84,
       author = {{Gallagher}, III, J.~S. and {Hunter}, D.~A. and {Tutukov}, A.~V.},
        title = "{Star formation histories of irregular galaxies.}",
      journal = {\apj},
         year = 1984,
        month = sep,
       volume = {284},
        pages = {544-556},
          doi = {10.1086/162437},
       adsurl = {https://ui.adsabs.harvard.edu/abs/1984ApJ...284..544G}
}

@ARTICLE{Tolstoy09,
       author = {{Tolstoy}, Eline and {Hill}, Vanessa and {Tosi}, Monica},
        title = "{Star-Formation Histories, Abundances, and Kinematics of Dwarf Galaxies in the Local Group}",
      journal = {\araa},
         year = 2009,
        month = sep,
       volume = {47},
       number = {1},
        pages = {371-425},
          doi = {10.1146/annurev-astro-082708-101650},
archivePrefix = {arXiv},
       eprint = {0904.4505},
 primaryClass = {astro-ph.CO},
       adsurl = {https://ui.adsabs.harvard.edu/abs/2009ARA&A..47..371T}
}

@ARTICLE{Geha12,
       author = {{Geha}, M. and {Blanton}, M.~R. and {Yan}, R. and {Tinker}, J.~L.},
        title = "{A Stellar Mass Threshold for Quenching of Field Galaxies}",
      journal = {\apj},
         year = 2012,
        month = sep,
       volume = {757},
       number = {1},
          eid = {85},
        pages = {85},
          doi = {10.1088/0004-637X/757/1/85},
archivePrefix = {arXiv},
       eprint = {1206.3573},
 primaryClass = {astro-ph.CO},
       adsurl = {https://ui.adsabs.harvard.edu/abs/2012ApJ...757...85G}
}

@ARTICLE{Geha24,
       author = {{Geha}, Marla and {Mao}, Yao-Yuan and {Wechsler}, Risa H. and {Asali}, Yasmeen and {Kado-Fong}, Erin and {Kallivayalil}, Nitya and {Nadler}, Ethan O. and {Tollerud}, Erik J. and {Weiner}, Benjamin and {de los Reyes}, Mithi A.~C. and {Wang}, Yunchong and {Wu}, John F.},
        title = "{The SAGA Survey. IV. The Star Formation Properties of 101 Satellite Systems around Milky Way{\textendash}mass Galaxies}",
      journal = {\apj},
         year = 2024,
        month = nov,
       volume = {976},
       number = {1},
          eid = {118},
        pages = {118},
          doi = {10.3847/1538-4357/ad61e7},
archivePrefix = {arXiv},
       eprint = {2404.14499},
 primaryClass = {astro-ph.GA},
       adsurl = {https://ui.adsabs.harvard.edu/abs/2024ApJ...976..118G}
}

@ARTICLE{Sparre17,
       author = {{Sparre}, Martin and {Hayward}, Christopher C. and {Feldmann}, Robert and {Faucher-Gigu{\`e}re}, Claude-Andr{\'e} and {Muratov}, Alexander L. and {Kere{\v{s}}}, Du{\v{s}}an and {Hopkins}, Philip F.},
        title = "{(Star)bursts of FIRE: observational signatures of bursty star formation in galaxies}",
      journal = {\mnras},
         year = 2017,
        month = apr,
       volume = {466},
       number = {1},
        pages = {88-104},
          doi = {10.1093/mnras/stw3011},
archivePrefix = {arXiv},
       eprint = {1510.03869},
 primaryClass = {astro-ph.GA},
       adsurl = {https://ui.adsabs.harvard.edu/abs/2017MNRAS.466...88S}
}

@ARTICLE{Renzini15,
       author = {{Renzini}, Alvio and {Peng}, Ying-jie},
        title = "{An Objective Definition for the Main Sequence of Star-forming Galaxies}",
      journal = {\apjl},
         year = 2015,
        month = mar,
       volume = {801},
       number = {2},
          eid = {L29},
        pages = {L29},
          doi = {10.1088/2041-8205/801/2/L29},
archivePrefix = {arXiv},
       eprint = {1502.01027},
 primaryClass = {astro-ph.GA},
       adsurl = {https://ui.adsabs.harvard.edu/abs/2015ApJ...801L..29R}
}

@ARTICLE{Kurczynski16,
       author = {{Kurczynski}, Peter and {Gawiser}, Eric and {Acquaviva}, Viviana and {Bell}, Eric F. and {Dekel}, Avishai and {de Mello}, Duilia F. and {Ferguson}, Henry C. and {Gardner}, Jonathan P. and {Grogin}, Norman A. and {Guo}, Yicheng and {Hopkins}, Philip F. and {Koekemoer}, Anton M. and {Koo}, David C. and {Lee}, Seong-Kook and {Mobasher}, Bahram and {Primack}, Joel R. and {Rafelski}, Marc and {Soto}, Emmaris and {Teplitz}, Harry I.},
        title = "{Evolution of Intrinsic Scatter in the SFR-Stellar Mass Correlation at 0.5 < z < 3}",
      journal = {\apjl},
         year = 2016,
        month = mar,
       volume = {820},
       number = {1},
          eid = {L1},
        pages = {L1},
          doi = {10.3847/2041-8205/820/1/L1},
archivePrefix = {arXiv},
       eprint = {1602.03909},
 primaryClass = {astro-ph.GA},
       adsurl = {https://ui.adsabs.harvard.edu/abs/2016ApJ...820L...1K}
}

@ARTICLE{Koller24,
       author = {{Koller}, M. and {Ziegler}, B. and {Ciocan}, B.~I. and {Thater}, S. and {Mendel}, J.~T. and {Wisnioski}, E. and {Battisti}, A.~J. and {Harborne}, K.~E. and {Foster}, C. and {Lagos}, C. and {Croom}, S.~M. and {Grasha}, K. and {Papaderos}, P. and {Remus}, R.~S. and {Sharma}, G. and {Sweet}, S.~M. and {Valenzuela}, L.~M. and {van de Ven}, G. and {Zafar}, T.},
        title = "{The MAGPI survey: The interdependence of the mass, star formation rate, and metallicity in galaxies at z {\ensuremath{\sim}} 0.3}",
      journal = {\aap},
         year = 2024,
        month = sep,
       volume = {689},
          eid = {A315},
        pages = {A315},
          doi = {10.1051/0004-6361/202450715},
archivePrefix = {arXiv},
       eprint = {2406.20017},
 primaryClass = {astro-ph.GA},
       adsurl = {https://ui.adsabs.harvard.edu/abs/2024A&A...689A.315K}
}

\end{document}